\documentclass[reprint, superscriptaddress, amsmath,amssymb, prl, showpacs, altaffilsymbol, longbibliography]{revtex4-1}

\usepackage{graphicx}  
\usepackage{braket}
\usepackage{color}
\usepackage{mathtools}

\begin{document}

\title{A programmable two-qubit quantum processor in silicon}
\author{T.~F.~Watson}
\email{tfwatson15@gmail.com}
\affiliation{QuTech and the Kavli Institute of Nanoscience, Delft University of Technology, 5046, 2600 GA Delft, Netherlands}
\author{S.~G.~J.~Philips}
\affiliation{QuTech and the Kavli Institute of Nanoscience, Delft University of Technology, 5046, 2600 GA Delft, Netherlands}
\author{E.~Kawakami}
\affiliation{QuTech and the Kavli Institute of Nanoscience, Delft University of Technology, 5046, 2600 GA Delft, Netherlands}
\author{D.~R.~Ward}
\affiliation{University of Wisconsin-Madison, Madison, WI 53706, USA}
\author{P.~Scarlino}
\affiliation{QuTech and the Kavli Institute of Nanoscience, Delft University of Technology, 5046, 2600 GA Delft, Netherlands}
\author{M.~Veldhorst}
\affiliation{QuTech and the Kavli Institute of Nanoscience, Delft University of Technology, 5046, 2600 GA Delft, Netherlands}
\author{D.~E.~Savage}
\affiliation{University of Wisconsin-Madison, Madison, WI 53706, USA}
\author{M.~G.~Lagally}
\affiliation{University of Wisconsin-Madison, Madison, WI 53706, USA}
\author{Mark~Friesen}
\affiliation{University of Wisconsin-Madison, Madison, WI 53706, USA}
\author{S.~N.~Coppersmith}
\affiliation{University of Wisconsin-Madison, Madison, WI 53706, USA}
\author{M.~A.~Eriksson}
\affiliation{University of Wisconsin-Madison, Madison, WI 53706, USA}
\author{L.~M.~K.~Vandersypen}
\email{L.M.K.Vandersypen@tudelft.nl}
\affiliation{QuTech and the Kavli Institute of Nanoscience, Delft University of Technology, 5046, 2600 GA Delft, Netherlands}
\date{\today}%
\maketitle
\textbf{With qubit measurement and control fidelities above the threshold of fault-tolerance, much attention is moving towards the daunting task of scaling up the number of physical qubits to the large numbers needed for fault tolerant quantum computing \cite{Barends2014, Debnath2016}. Here, quantum dot based spin qubits may offer significant advantages due to their potential for high densities, all-electrical operation, and integration onto an industrial platform \cite{Loss1998, Maurand2016, Vandersypen2017}. In this system, the initialisation, readout, single- and two-qubit gates have been demonstrated in various qubit representations \cite{Shulman2012,Kim2014, Veldhorst2015, Medford2013b}.  However, as seen with other small scale quantum computer demonstrations \cite{Vandersypen2001, DiCarlo2009, Gulde2003, Sar2012}, combining these elements leads to new challenges involving qubit crosstalk, state leakage, calibration, and control hardware. Here we show that these challenges can be overcome by demonstrating a programmable two-qubit quantum processor in silicon by performing both the Deutsch-Josza and the Grover search algorithms. In addition, we characterise the entanglement in our processor through quantum state tomography of Bell states measuring state fidelities between 85-89$\%$ and concurrences between 73-82$\%$. These results pave the way for larger scale quantum computers using spins confined to quantum dots.}

\begin{figure*}[ht]
\includegraphics[scale=0.95]{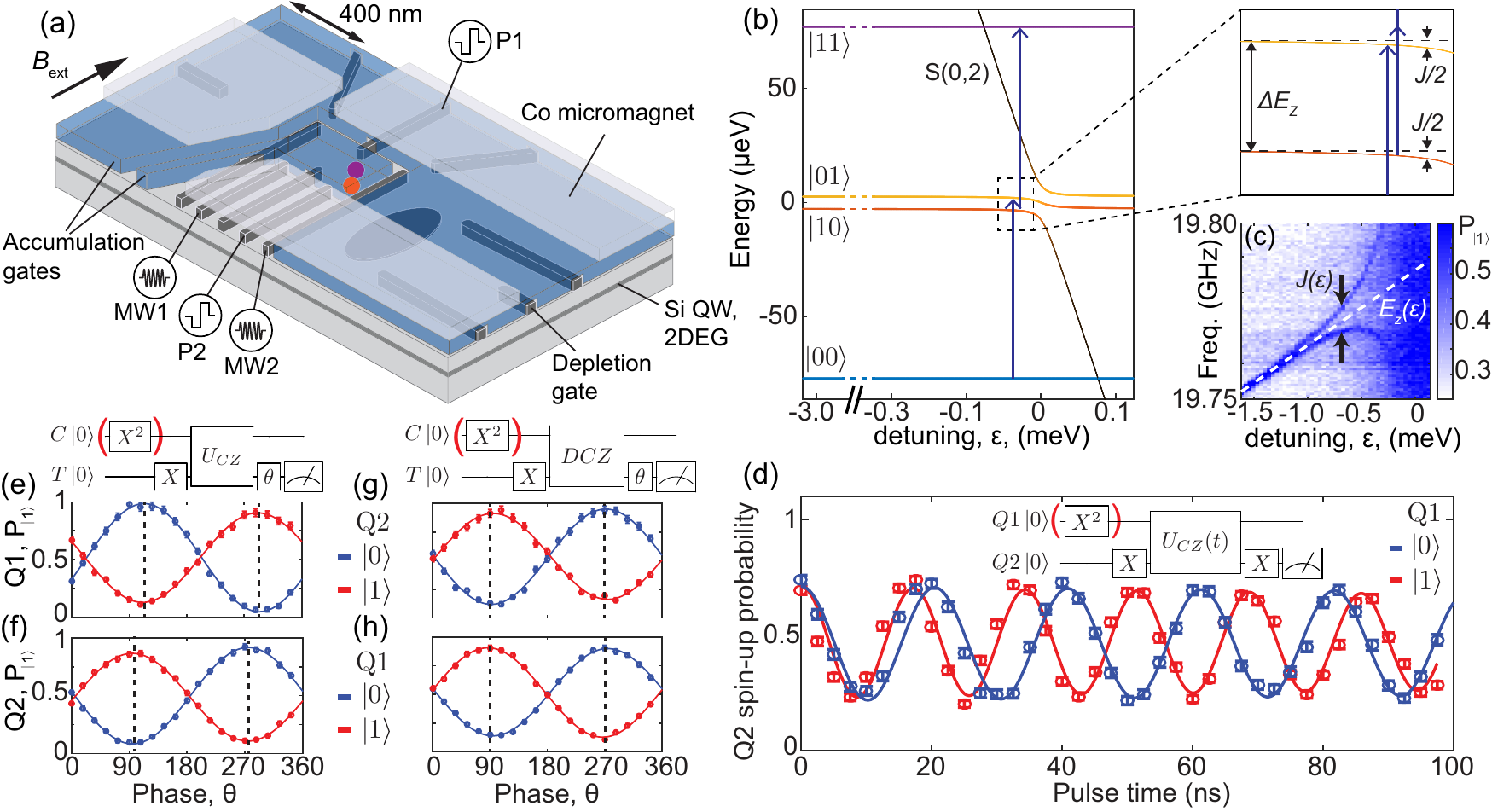}
\caption{{\bf Two-qubit quantum processor in silicon.} (a) Schematic of a Si/SiGe double quantum dot device showing the estimated position of quantum dots D1 (purple circle) and D2 (orange circle) used to confine two electron spin qubits Q1 and Q2, respectively. Both quantum dots were formed on the right side of the device to achieve an interdot tunnel coupling suitable for two-qubit gates. The position of the dots was realised through the tuning of the numerous electrostatic gates but was most likely helped by disorder in the Si/SiGe heterostructure. The ellipse shows the position of the QD sensor used for spin readout. Microwave signals MW1 and MW2 are used to perform EDSR on Q1 and Q2, respectively, while voltage pulses are applied to plunger gates P1 and P2 for qubit manipulation and readout. (b) Energy level diagram of two electron spins in a double quantum dot as a function of the detuning energy, $\epsilon$, between the (1,1) and (0,2) charge states. (c) Microwave spectroscopy of Q2 versus detuning energy after initialisation of Q1  to $ (\ket{0} + \ket{1})/\sqrt{2}$. The detuning voltage was converted to energy using a lever arm of $\alpha = 0.09e$ (see Extended Data Fig.~5). The map shows that Q2 has two different resonant frequencies (blue arrows in (b)) depending on the spin state of Q1, which are separated by the exchange energy, $J$. (d) The spin-up probability of Q2 as a function of the detuning pulse duration in a Ramsey sequence with the control Q1 initialised to spin-down (blue curve) and spin-up (red curve). (e-f) Calibration of the $\hat{z}$ rotations on Q1 and Q2 needed to form the $CZ_{ij}$ gates are performed by using a Ramsey sequence and varying the phase of the last $\pi/2$ pulse. Here the spin-up probability has been normalised to remove initialisation and readout errors and the exchange energy is $J/h = 10$~MHz. (g,h) A decoupled version of the CZ gate removes the unconditional $\hat{z}$ rotations due to the detuning dependence on $E_Z (\epsilon)$. Consequently, the required $\hat{z}$ rotations to form the $CZ_{ij}$ gates (dashed black lines) are always at $90^\circ$ and $270^\circ$, simplifying calibration. All error bars are $1\sigma$ from the mean calculated from a Monte Carlo estimation (see methods).}
\label{Figure1}
\end{figure*}

Solid-state approaches to quantum computing are challenging to realise due to unwanted interactions between the qubit and the host material. For quantum dot based qubits, charge and nuclear spin noise are the dominant sources of decoherence and gate errors.  While some of these effects can be cancelled out by using dynamical decoupling \cite{Bluhm2011} or decoherence-free subspaces \cite{Petta2005, Medford2013b}, there has also been  significant progress in reducing these noise sources through growing better oxides and heterostructures \cite{Zwanenburg2013} and moving to silicon (Si) due to its naturally low abundance of nuclear spin isotopes which can be  removed through isotopic purification \cite{Tyryshkin2012}. These material developments have dramatically extended qubit coherence times enabling single-qubit gate fidelities above 99$\%$ \cite{Veldhorst2014,Muhonen2014, Kawakami2016, Yoneda2017} and recently resulted in the demonstration of a controlled phase (CZ) gate between two single electron spin qubits in a silicon metal-oxide-semiconductor (Si-MOS) device \cite{Veldhorst2015}. Here, we show that with two single electron spin qubits in a natural silicon/silicon-germanium (Si/SiGe) double quantum dot (DQD), we can combine initialisation, readout, single- and two-qubit gates to form a programmable quantum processor in silicon that can perform simple quantum algorithms. 

A schematic of the two-qubit quantum processor is shown in Fig.~1(a). The device is similar to that described in \cite{Kawakami2014} except for an additional micromagnet. A two-dimensional electron gas (2DEG) is formed in the natural Si quantum well of a SiGe heterostructure using two accumulation gates. The DQD is defined in the 2DEG by applying negative voltages to the depletion gates with the estimated position of the first (D1) and second (D2) quantum dot shown by the purple and orange circle, respectively. The two qubits, Q1 and Q2, are defined by applying a finite magnetic field of $B_{ext} = 617$~mT and using the Zeeman-split spin-down~$\ket{0}$ and spin-up~$\ket{1}$ states of single electrons respectively confined in D1 and D2. The initialisation and readout of Q2 is performed by spin-selective tunnelling to a reservoir \cite{Elzerman2004} while Q1 is initialised at a spin relaxation hotspot \cite{Srinivasa2013} and measured via Q2 using a controlled rotation (CROT). The complete measurement sequence and setup are described in Extended Data Fig.~1,2 where we achieve initialisation and  readout fidelities of $F_{I1}>99\%$, $F_{I2}>99\%$,  $F_{m1} = 73\%$, and $F_{m2} = 81\%$ (see methods). 

The coherent individual control of both qubits is achieved by patterning three cobalt micromagnets on top of the device (see Fig.~1(a)). These micromagnets provide a magnetic field gradient with a component that is perpendicular to the external magnetic field for electric dipole spin resonance (EDSR) \cite{Pioro-Ladriere2008}. Furthermore, the field gradient across the two dots results in qubit frequencies that are well separated   ($f_{Q1} = 18.4$~GHz, $f_{Q2} = 19.7$~GHz), allowing the qubits to be addressed independently. For both qubits, we achieve Rabi frequencies of $f_R = w_R/2\pi = 2$~MHz and perform single qubit X and Y gates by using vector modulation of the microwave (MW) drive signals. Here, we define an X (Y) gate to be a $\pi/2$ rotation around $\hat{x}$ ($\hat{y}$) and henceforth define a $\pi$ rotation to be $X^2$ ($Y^2$). We measure the qubit properties of Q1 (Q2) in the (1,1) regime (where $(m,n)$ denotes a configuration with $m$ electrons in D1 and $n$ electrons in D2) to be $T_1 > 50$~ms ($3.7\pm 0.5$~ms), $T_2^* = 1.0\pm0.1$~$\mu$s ($0.6\pm0.1$~$\mu$s), $T_{2Hahn} = 19\pm3$~$\mu$s ($7\pm1$~$\mu$s) (see Extended Data Fig.~3). Using single qubit randomised benchmarking \cite{Knill2008, Kawakami2016} we find an average Clifford gate fidelity of 98.8$\%$ for Q1 and 98.0$\%$ for Q2 (see Extended Data Fig.~4) which are close to the fault tolerant error threshold for surface codes \cite{Fowler2012}. 

Universal quantum computing requires the implementation of both single- and two-qubit gates. In this quantum processor we implement a two-qubit controlled-phase (CZ) gate \cite{Meunier2011, Veldhorst2015}. This gate can be understood by considering the energy level diagram for two electron spins in a double quantum dot, shown in Fig.~1(b), in the regime where the Zeeman energy difference is comparable to the interdot tunnel coupling, $\delta E_Z \sim t_c$. The energies of the two-spin states ($\ket{00}$, $\ket{01}$, $\ket{10}$, $\ket{11}$) in the (1,1) charge regime and the singlet ground state in the (0,2) charge regime are plotted as a function of the detuning, $\epsilon$. Here, detuning describes the energy difference between the (1,1) and (0,2) charge states of the DQD, controlled with the voltage applied to gate P1 (see Extended Data Fig.~2). The anticrossing between the S(0,2) and the antiparallel  $\ket{01}$ and $\ket{10}$ states causes the energy of the antiparallel states to decrease by $J(\epsilon)/2$ as the detuning is decreased (see Fig.~1(b)), where $J(\epsilon)$ is the exchange coupling between the two electron spins. 
 
The energy structure of the two-electron system can be probed by performing MW spectroscopy as a function of detuning as shown in Fig.~1(c). At negative detuning, the resonance frequency (Zeeman energy) increases linearly (dashed line) due to the electron wavefunction moving in the magnetic field gradient. At more positive detuning closer to the (0,2) regime, the exchange energy is significant compared to the linewidth of the resonance  $J/h>\omega_R$, resulting in two clear resonances. Applying a $\pi$ pulse at one of these frequencies results in a CROT gate which is used to perform the projective measurement of Q1 via the readout of Q2 (see Extended Data Fig.~6). 

The CZ gate is implemented by applying a detuning pulse  for a fixed amount of time, $t$, which shifts the energy of the antiparallel states. Throughout the pulse, we stay in the regime where $J(\epsilon) \ll \Delta E_z$, so the energy eigenstates of the system are still the two-spin product states and the two-qubit interaction can be approximated by an Ising Hamiltonian, leading to the following unitary operation, 
\begin{eqnarray}
U_{CZ} (t) = 
Z_1(\theta_1) Z_2(\theta_2)
\begin{pmatrix}
1 & 0 & 0 & 0 \\
0 &e^{i J(\epsilon) t/2\hbar}& 0 & 0\\
0 & 0 & e^{i J(\epsilon) t/2\hbar} & 0\\
0 &0 & 0 & 1
\end{pmatrix},
\end{eqnarray} 
where the basis states are $\ket{00}$, $\ket{01}$, $\ket{10}$, and, $\ket{11}$, and $Z_1(\theta_1)$ and $Z_2(\theta_2)$ are rotations around $\hat{z}$ caused by the change in the Zeeman energy of the qubits due to the magnetic field gradient. The CZ gate is advantageous over the CROT as it is faster and less time is spent at low detuning, where the qubits are more sensitive to charge noise. In addition, we observed that performing the CROT with EDSR can lead to state leakage into the S(0,2) state, seen in Fig.~1(c) by the increase in background dark counts near $\epsilon =0
$. The CZ gate is demonstrated in Fig.~1(d); the duration of a CZ voltage pulse between two X gates on Q2 in a Ramsey experiment is varied, showing that the frequency of the $\hat{z}$ rotation on Q2 is conditional on the spin state of Q1. The processor's primitive two-qubit gates, $CZ_{ij} \ket{m,n} = (-1)^{\delta(i,m)\delta(j,n)} \ket{m,n}$ for $i,j,m,n \in \{0,1\}$, are constructed by applying the CZ gate for a time $t = \pi \hbar/J$ followed by $\hat{z}$ rotations on Q1 and Q2, $CZ_{ij} = Z_1((-1)^j\pi/2-\theta_1)Z_2((-1)^i\pi/2-\theta_2) U_{CZ}(\pi \hbar/J)$. Rather than physically performing the $\hat{z}$ rotations, we use a software reference frame change where we incorporate the rotation angle $\theta_1$ and $\theta_2$ into the phase of any subsequent MW pulses \cite{Vandersypen2001}.  

Combining single- and two-qubit gates together with initialisation and readout, we demonstrate a programmable processor --- where we can program arbitrary sequences for the two-qubit chip to execute within the coherence times of the qubits. To achieve this, a number of challenges needed to be overcome. The device had to be further tuned so that during single-qubit gates the exchange coupling was low, $J_{\mathit{off}}/h = 0.27$~MHz (see Extended Data Fig.~7),  compared to our single-qubit gate times ($\sim 2$~MHz) and two-qubit gate times ($\sim 6 - 10$~MHz). Tuning was also required to raise the energy of low-lying valley-excited states to prevent them from being populated during initialisation \cite{Kawakami2014}. Furthermore, we observed that applying MW pulses on Q1 shifts the resonance frequency of Q2 by $\sim 2$~MHz. We rule out the AC Stark shift, effects from coupling between the spins, and heating effects as possible explanations  but find the quantum dot properties affect the frequency shift (see Supplementary information S1). While the origin of the shift is unknown, we keep the resonance frequency of Q2 fixed during single-qubit gates by applying an off-resonant pulse ($30$~MHz) to Q1 if Q1 is idle.

Before running sequences on the quantum processor, all gates need to be properly calibrated. The single-qubit X and Y gates were calibrated using both a Ramsey sequence and the AllXY calibration sequence to determine the qubit resonance frequency and the power needed to perform a $\pi/2$ gate (see Supplementary information S2). To calibrate the $CZ_{ij}$ gates we performed the Ramsey sequence in Fig.~1(e) and varied the phase of the last $\pi/2$ gate. Fig.~1(e) shows the results of this measurement where Q1 is the target qubit and the control qubit Q2 is either prepared in $\ket{0}$ (blue curve) or $\ket{1}$ (red curve).  The duration of the CZ gate is calibrated so that the blue and red curve are 180$^\circ$ out of phase. These measurements also determine the $\hat{z}$ rotation on Q1 needed to form $CZ_{ij}$, which corresponds to the phase of the last $\pi/2$ gate which either maximises or minimises the Q2 spin-up probability (dashed lines in Fig.~1(e)). The $\hat{z}$ rotation needed for Q2 is calibrated by performing a similar measurement, where the roles of Q1 and Q2 are switched (Fig.~1(f)).

\begin{figure}[t]
\begin{center}
\includegraphics[scale=0.9]{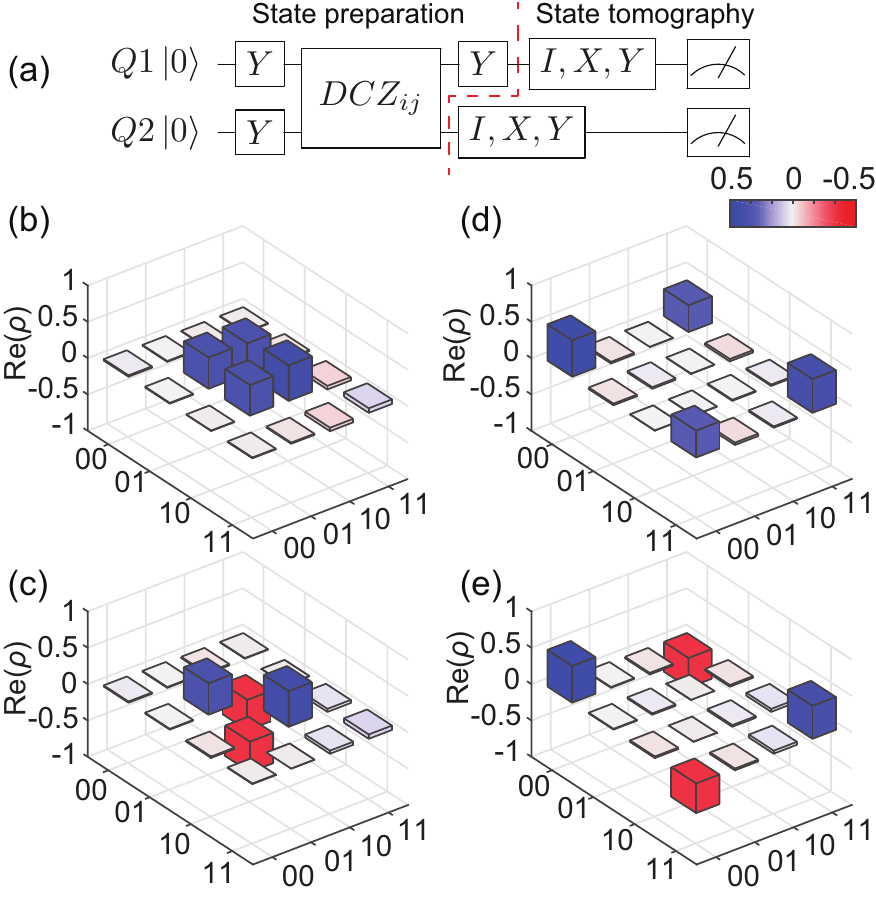}
\caption{{\bf Preparation of the Bell states and two-qubit entanglement in silicon.}
 (a) The quantum circuit used to prepare the Bell states and perform quantum state tomography. (b-e) The real component of the reconstructed density matrices using a maximum likelihood estimation for the four Bell states (b) $\Psi^+ = (\ket{01} + \ket{10})/\sqrt{2}$, (c) $\Psi^- = (\ket{01} - \ket{10})/\sqrt{2}$, (d) $\Phi^+ = (\ket{00} + \ket{11})/\sqrt{2}$, (e) $\Phi^- =  (\ket{00} - \ket{11})/\sqrt{2}$. The imaginary components of the density matrices are $<0.08$ for all elements (see supplementary information S3). We measure state fidelities of $F_{\Psi^+} = 0.88 \pm 0.02$, $F_{\Psi^-} = 0.88 \pm 0.02$,  $F_{\Phi^+} = 0.85 \pm 0.02$, $F_{\Phi^-} = 0.89 \pm 0.02$ and concurrences of $c_{\Psi^+} = 0.80 \pm 0.03$, $c_{\Psi^-} = 0.82\pm0.03$, $c_{\Phi^+}  = 0.73 \pm 0.03$, $c_{\Phi^-}  = 0.79 \pm 0.03$. All errors are $1\sigma$ from the mean.}
\label{Figure2}
\end{center}
\end{figure}

\begin{figure*}[ht]
\includegraphics[scale=0.95]{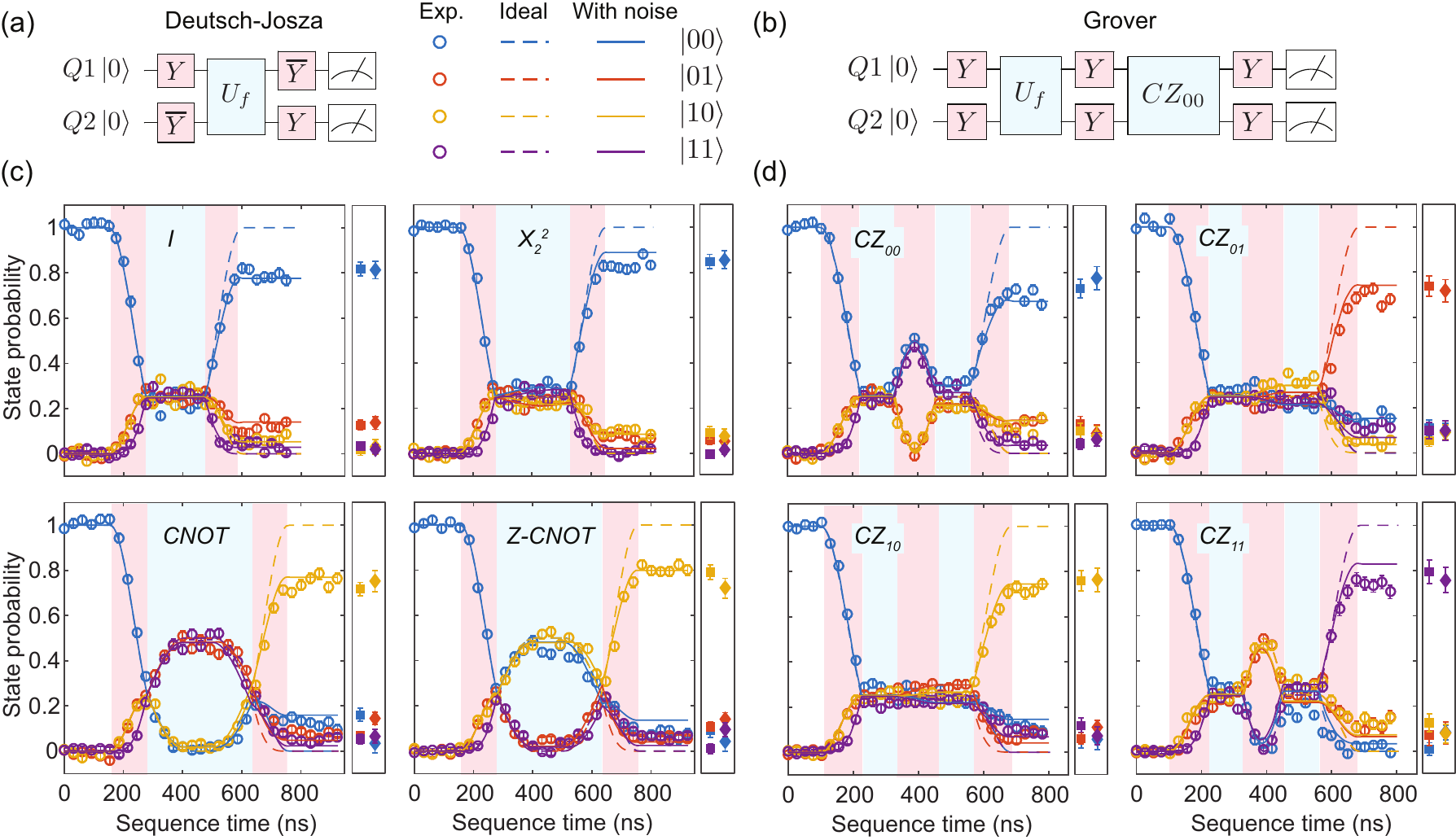}
\caption{{\bf Two-qubit quantum algorithms in silicon.} 
(a,b) The quantum circuits for the (a) Deutsch-Josza algorithm and (b) Grover search algorithm for two qubits. (c,d) Two-spin probabilities as a function of time throughout the sequence during the (c) Deutsch-Josza algorithm  and the (d) Grover search algorithm for each of four possible functions. Each point corresponds to 4000 repetitions and has been normalised to remove readout errors. The dash lines are the simulated ideal cases while the solid lines are the simulated results where decoherence is introduced by including quasistatic nuclear spin noise and charge noise ($\sigma_\epsilon$ = 11~$\mu$eV). For both algorithms, the square data points show the final results of the algorithms where all four functions are evaluated in the same measurement run with identical calibration. The diamonds show the result of both algorithms when using the decoupled CZ gate showing similar performance. For the Deutsch-Josza  algorithm the identity is implemented as either a 200~ns wait (circle and square data points) or as $I = X_1^4 X_2^4$ (diamond data points).  All error bars are $1\sigma$ from the mean.}
\label{Figure3}
\end{figure*}

The $\hat{z}$ rotations in Eq.~1 can be eliminated by using a decoupled CZ gate $DCZ = U_{CZ}(\pi\hbar/2J)$ $X^2_1 X^2_2 U_{CZ}(\pi\hbar/2J)$ which incorporates refocusing pulses and can be used to perform  $DCZ_{ij} = X_1^2 X_2^2 CZ_{ij}$ $= Z_1((-1)^j\pi/2) Z_2 ((-1)^i\pi/2) DCZ$. This is demonstrated in the Ramsey experiment in Fig.~1(g,h), where the minimum and maximum spin-up probabilities occur at a phase of either 90$^\circ$ or 270$^\circ$. In addition to removing the need to calibrate the required $\hat{z}$ rotations, this gate is advantageous as it cancels out the effect of low frequency noise that couples to the spins via $\sigma_Z \otimes I$ and $I \otimes \sigma_Z$ terms during the gate.

After proper calibration, we can characterise entanglement in our quantum processor by preparing Bell states and reconstructing the two-qubit density matrix using quantum state tomography. The quantum circuit for the experiment is shown in Fig.~2(a). The Bell states are prepared using a combination of single-qubit gates and the decoupled two-qubit $DCZ_{ij}$ gates. The density matrix is reconstructed by measuring two-spin probabilities for the 9 combinations of 3 different measurement bases (x,y,z) with 10,000 repetitions (see methods). In our readout scheme the states are projected into the z-basis while measurements in the other bases are achieved by performing X and Y pre-rotations. Due to the time needed to perform these measurements ($\sim 2$~hrs) the frequency of the qubits  was calibrated after every 100 repetitions. The real components of the reconstructed density matrices of the four Bell states ($1/\sqrt{2} (\ket{00} \pm \ket{11})$, $1/\sqrt{2} (\ket{01} \pm \ket{10})$) are shown in Fig.~2(b-e). The state fidelities, $F = \bra{\psi} \rho \ket{\psi}$, between these density matrices and the target Bell states range between 85-89$\%$ and the concurrences range between 73-82$\%$, demonstrating entanglement. A parallel experiment reported a 78\% Bell state fidelity \cite{Zajac2018}.

To test the programmability of the two-qubit quantum processor we perform the Deutsch-Josza \cite{Deutsch1992} and the Grover search \cite{Grover1997} quantum algorithms.  The Deutsch-Josza algorithm determines whether a function is constant $(f_1(0) = f_1(1) =0$ or $f_2(0) = f_2(1) =1$) or balanced ($f_3(0) =0$, $f_3(1) =1$ or $f_4(0) =1$, $f_4(1) =0$). These four functions are mapped onto the following unitary operators, $U_{f1} = I$, $U_{f2} = X_2^2$, $U_{f3} = CNOT = Y_2 CZ_{11} \overline{Y}_2 $, $U_{f4} = Z\textnormal{-}CNOT =\overline{Y}_2 CZ_{00} Y_2$ where the overline denotes a negative rotation. For both the controlled NOT (CNOT) and the zero-controlled NOT (Z-CNOT) the target qubit is Q2. At the end of the sequence  the input qubit (Q1) will  be either $\ket{0}$ or $\ket{1}$ for the constant and balanced functions, respectively. Grover's search algorithm provides an optimal method for finding the unique input value $x_0$ of a function $f(x)$  that gives $f(x_0) =1$ where $f(x) = 0$ for all other values of $x$. In the two-qubit version of this algorithm there are four input values, $x \in \{00, 01, 10, 11\}$, resulting in four possible functions, $f_{ij}(x)$ where $i,j \in \{0,1\}$. These functions are mapped onto the unitary operators, $CZ_{ij} \ket{x} = (-1)^{f_{ij}(x)}\ket{x}$, which mark the input state with a negative phase if $f_{ij}(x) =1$.  The algorithm finds the state that has been marked and outputs it at the end of the sequence.

Fig.~3 shows the measured two-spin probabilities as a function of time during the algorithms for each function. The experimental results (circles) are in good agreement with the simulated ideal cases (dashed lines). Although a number of repetitions are needed due to gate and readout errors, the algorithms are successful at determining the balanced and constant functions and finding the marked state in the oracle functions. The square data points are taken shortly after calibration and are in line with the circle data points, indicating that calibrations remain stable throughout the hour of data collection for the main panels. The diamond data points show the outcome of the algorithms using the decoupled CZ gate. In most cases, the diamond data points also give similar values to the circles, which means that the decoupled CZ gate does not improve the final result. This suggests that low-frequency single-qubit noise during the CZ gate is not dominant. The substantial difference between Hahn echo and Ramsey decay times still points at significant low-frequency noise. Single-qubit low-frequency noise, whether from nuclear spins or charge noise, reduces single-qubit coherence in particular during wait and idle times in the algorithms. Additionally charge noise affects the coupling strength $J$ during the CZ gates. Numerical simulations (solid lines in Fig.~3(c,d) and Extended Data Fig.~10) show that quasi-static nuclear spin noise and charge noise can reproduce most features seen in the two-qubit algorithm data (see Methods). Smaller error contributions include residual coupling during single-qubit operations and miscalibrations.

Significant improvements could be made in the performance of the processor by using isotopically purified $^{28}$Si \cite{Muhonen2014, Veldhorst2014, Yoneda2017}, which would increase the qubit coherence times. Furthermore, recent experiments have shown that symmetrically operating an exchange gate by pulsing the tunnel coupling rather than detuning leads to a gate which is less sensitive to charge noise, significantly improving fidelities \cite{Reed2016, Martins2016}. With these modest improvements combined with more reproducible and scalable device structures, quantum computers with multiple qubits and fidelities above the fault tolerant threshold should be realisable.

\noindent
{\bf Acknowledgements:} 
Research was sponsored by the Army Research Office (ARO), and was accomplished under Grant Numbers W911NF-17-1-0274 and W911NF-12-1-0607. The views and conclusions contained in this document are those of the authors and should not be interpreted as representing the official policies, either expressed or implied, of the Army Research Office (ARO), or the U.S. Government. The U.S. Government is authorized to reproduce and distribute reprints for Government purposes notwithstanding any copyright notation herein.  Development and maintenance of the growth facilities used for fabricating samples is supported by DOE (DE-FG02-03ER46028). We acknowledge the use of facilities supported by NSF through the UW-Madison MRSEC (DMR-1121288). E.K. was supported by a fellowship from the Nakajima Foundation. We acknowledge financial support by the Marie Sk\l{}odowska-Curie actions - Nanoscale solid-state spin systems in emerging quantum technologies - Spin-NANO, grant agreement number 676108. The authors acknowledge useful discussion with S.~Dobrovitski, C.~Dickel, A.~Rol, J.~P. Dehollain, Z.~Ramlakhan, members of the Vandersypen group, and technical assistance from R.~Schouten, R.~Vermeulen, M.~Tiggelman, M.~Ammerlaan, J.~Haanstra, R.~Roeleveld, O.~Benningshof.  
\newline
{\bf Author Contributions:} T.F.W performed the experiment with help from E.K. and P.S., T.F.W. and S.G.J.P. analysed the data, S.G.J.P performed the simulations of the algorithms, T.F.W, S.G.J.P, E.K., P. S., M.V., M.F., S.N.C., M.A.E. and L.M.K.V. contributed to the interpretation of the data and commented on the manuscript, D.R.W fabricated the device, D.E.S. and M.G.L. grew the Si/SiGe heterostructure, T.F.W. wrote the manuscript (S.G.J.P wrote parts of the methods), L.M.K.V. conceived and supervised the project.

\bibliography{references2017}

\setcounter{equation}{0}
\setcounter{figure}{0}
\setcounter{table}{0}
\makeatletter
\renewcommand{\theequation}{\arabic{equation}}
\renewcommand{\thefigure}{\arabic{figure}}
\renewcommand{\figurename}{Extended Data Fig.}

\section{Methods}
\noindent{\bf Estimation of initialisation and readout errors for Q1 and Q2.} The initialisation and readout procedures for Q1 and Q2 are described in the Extended Data Fig.~2.  The initialisation and readout fidelities of Q2 were extracted by performing the following three experiments and measuring the resulting spin-up probabilities ($P_1$, $P_2$, $P_3$): (i) Initialise Q2 and wait $7T_1$. (ii) Initialise Q2. (iii)  Initialise and perform a $\pi$ rotation on Q2. These three spin-up probabilities are related to the initialisation fidelity ($\gamma_2$) and the spin-up and spin-down readout fidelities ($F_{\ket{0},2}$, $F_{\ket{1},2}$)  by, 
\begin{eqnarray}
P_1 &=& 1-F_{\ket{0},2}, \\
P_2 &=& F_{\ket{1},2} (1-\gamma_2) + (1-F_{\ket{0},2})\gamma_2, \\
P_3/P_{\pi2} &=& F_{\ket{1},2} (\gamma_2) + (1-F_{\ket{0},2})(1-\gamma_2),
\label{initfidelity}
\end{eqnarray}
where $P_{\pi2}$ is the expected probability to be in the up state after the application of the $\pi$ pulse for Q2, which is determined as described below. In Eq.~\ref{initfidelity} we assume that waiting $7T_1$ leads to 100\% initialisation and the measured spin-up counts are due to the readout infidelity. By solving these three equations we can extract the initialisation and readout fidelities. For Q1, we performed initialisation by pulsing to a spin relaxation hotspot (see Extended Data Fig.~5) for $500 T_1$ and therefore we assume the initialisation fidelity is $\sim$100\%. Consequently, the readout fidelities of Q1 were extracted by only performing experiments (ii) and (iii) above.  The readout and initialisation fidelities for Q1 (Q2)  during the state tomography experiments were estimated to be   $\gamma_1 >99\%$ ($\gamma_2 >99\%$), $F_{\ket{0},1} = 92\%$ ($F_{\ket{0},2} = 86\%$), and $F_{\ket{1},1} = 54\%$ ($F_{\ket{1},2} = 76\%$) where we used $P_{\pi1} = 98\%$ ($P_{\pi2} = 97\%$) based on simulations which include the dephasing time of the qubits (see below). The average measurement fidelity, $F_m = (F_{\ket{0}} + F_{\ket{1}})/2$, for Q1(Q2) is 73\% (81\%). These fidelities are mostly limited by the finite electron temperature $T_e \approx 130$~mK and the fast spin relaxation time of Q2 ($T_1 = 3.7$~ms) , which is most likely caused by a spin relaxation hotspot due to a similar valley splitting and Zeeman energy \cite{Yang2013}.

\noindent{\bf Removing readout errors from the measured two-spin probabilities.}
In the experiment the measured two-spin probabilities $P^M =(P^M_{\ket{00}}, P^M_{\ket{01}}, P^M_{\ket{10}}, P^M_{\ket{11}})^T$ include errors due to the  limited readout fidelity $F_{\ket{0}, i}$ and $F_{\ket{1}, i}$, of a spin down $\ket{0}$ and spin up $\ket{1}$ electron for qubit $i$. To remove these readout errors to get the actual two-spin probabilities, $P =(P_{\ket{00}}, P_{\ket{01}}, P_{\ket{10}}, P_{\ket{11}})^T$, we use the following relationship, 
\begin{eqnarray}
P^M = (\hat{F}_1 \otimes \hat{F}_2) P
\end{eqnarray}
where, 
\begin{eqnarray}
\hat{F}_i = \begin{pmatrix}
F_{\ket{0},i} & 1-F_{\ket{1},i} \\
1-F_{\ket{0},i}& F_{\ket{1},i}
\end{pmatrix}.
\label{twospinconversion}
\end{eqnarray}

\noindent{\bf State tomography} The density matrix of a two-qubit state can be expressed as $\rho = \sum \limits^{16}_{i=1} c_i M_i$ where $M_i$ are 16 linearly independent measurement operators. The coefficients $c_i$ were calculated from the expectation values, $m_i$, of the measurement operators using a maximum likelihood estimation \cite{James2001, DiCarlo2009}. The expectation values were calculated by performing 16 combinations of  $I, X, Y, X^2$ prerotations on Q1 and Q2 and measuring the two-spin probabilities over 10,000 repetitions per measurement. The two-spin probabilities were converted to actual two-spin probabilities by removing readout errors using Eq.~\ref{twospinconversion}. For the calculation of the density matrices in Fig.~2 we only used the data from the $I, X, Y$ prerotations with the assumption that $I$ will give a more accurate estimation of the expectation values than $X^2$ due to gate infidelities. If we include the $X^2$ we achieve state fidelities between $80-84\%$ and concurrences between $67-71\%$ (see supplementary information S3).  In the analysis we assume the prerotations are perfect which is a reasonable approximation due to the high single-qubit Clifford gate fidelities $>98\%$ compared to the measured state fidelities $85-89\%$.  The state tomography experiment was performed in parallel with both the fidelity experiments described above and a Ramsey experiment used to actively calibrate the frequency. 

\noindent{\bf Error analysis.} Error analysis was performed throughout the manuscript using a Monte Carlo method by assuming a multinomial distribution for the measured two-spin probabilities and a binomial distribution for the probabilities ($P_1$, $P_2$, $P_3$) used to calculated the fidelities. Values from these distributions were randomly sampled and the procedures from above were followed. This was repeated 250 times to build up a final distributions which we use to determine the mean values and the standard deviation. 

\noindent{\bf Simulation of two electron spins in a double quantum dot.} In the simulation, we consider two electrons in two tunnel-coupled quantum dots where an external magnetic field $B_0$ is applied to both dots. In addition to this field, the two dots have different Zeeman energies due to the magnetic field gradient across the double quantum dot generated by micromagnets.  The Zeeman energy of Q1 (Q2) will be denoted as $B_1$ ($B_2$). The double dot system is modelled with the following Hamiltonian  \cite{DasSarma2011},
\begin{equation}
	\hat{H} = 
	\begin{pmatrix}
	-\beta & 0 & 0 & 0 & 0 & 0 \\
	0 &  -\Delta v & 0 & 0 & t & t \\
	0 &  0 & \Delta v & 0 & -t & -t \\
	0 &  0 & 0 & \beta & 0 & 0 \\
	0 &  t & -t & 0 & U_1 + \epsilon & 0 \\
	0 &  t & -t & 0 & 0 & U_2 - \epsilon\\
	\end{pmatrix},
\end{equation}
with the following states as the eigenbasis ($\ket{00}, \ket{01}, \ket{10}, \ket{11}, S(2,0), S(0,2)$). In this Hamiltonian, $\beta = \frac{B_1 + B_2}{2}$, $\Delta v = \frac{B_1 - B_2}{2}$, $\sqrt{2} t$ is the tunnel coupling between the (1,1) and (0,2)/(2,0) singlet states, and $U_i$ is the on-site charging energy of the i$^{th}$ quantum dot.
In order to study the phases of the qubits during control pulses, the Hamiltonian is transformed into a rotating frame using,
\begin{equation}\label{eq:rot_frame_trans}
	\widetilde{H} = VHV^\dagger + i \hbar (\partial_tV) V^\dagger, 
\end{equation}
where $V = e^{-i (B_1(\hat{\sigma}_z \otimes \hat{I}) + B_2(\hat{I} \otimes \hat{\sigma}_z))t}$ is the matrix that describes the unitary transformation where $\hbar = 1$. The transformed Hamiltonian is,
\begin{equation}
	\widetilde{H} =
	\begin{pmatrix}
	0 & 0 & 0 & 0 & 0 & 0 \\
	0 &  0 & 0 & 0 & t\:e^{i\Delta v t} & t\:e^{i\Delta v t} \\
	0 &  0 & 0 & 0 & -t\:e^{-i\Delta v t} & -t\:e^{-i\Delta v t} \\
	0 &  0 & 0 & 0 & 0 & 0 \\
	0 &  t\:e^{-i\Delta v t} & -t\:e^{i\Delta v t} & 0 & U_1 + \epsilon & 0 \\
	0 &  t\:e^{-i\Delta v t} & -t\:e^{i\Delta v t} & 0 & 0 & U_2 - \epsilon\\
	\end{pmatrix}.
\end{equation}
To model the single qubit gates during EDSR, we used the following Hamiltonian,
\begin{equation}
	\hat{H}_{mw} = \sum_k B_{mw,k}\cos{(\omega_k t + \phi_k)} [ \hat{\sigma}_x \otimes \hat{I}  + \hat{I} \otimes \hat{\sigma}_x],
\end{equation}
which assumes the same drive amplitude on each of the qubits. Here, $k$ represents the $k^{th}$ signal with an angular frequency $\omega_k$, phase $\phi_k$, and driving amplitude $B_{mw,k}$.  This Hamiltonian is transformed into the rotating frame using equation \ref{eq:rot_frame_trans} and the rotating wave approximation (RWA) can be made to remove the fast driving elements as the Rabi frequency is much smaller than the Larmor precession. This gives the following Hamiltonian, 
\begin{equation}
	\small
	\widetilde{H}_{mw} = \sum_k
	\begin{pmatrix}
	0 & \Omega_k e^{i\Delta\omega_1t} & \Omega_k e^{i\Delta\omega_2t} & 0 & 0 & 0 \\
	\Omega^*_k e^{-i\Delta\omega_1t} & 0 & 0 & \Omega_k e^{i\Delta\omega_2t} & 0 & 0\\
	\Omega^*_k e^{-i\Delta\omega_2t} & 0 & 0 & \Omega_k e^{i\Delta\omega_1t} & 0 & 0 \\
	0 & \Omega^*_k e^{-i\Delta\omega_2t} & \Omega^*_k e^{-i\Delta\omega_1t} & 0 & 0 & 0 \\
	0 & 0 & 0 & 0 & 0 & 0 \\
	0 & 0 & 0 & 0 & 0 & 0\\
	\end{pmatrix},
\end{equation}
where $\Omega_k$ is defined as $B_{MW,k}e^{i\phi_k}$, $\Omega_k^*$ is the complex conjugate of $\Omega$, and $\Delta\omega_k$ is defined as $\omega_{k} - \omega_{qubit_i}$.

The dynamics of the two qubit system can be described by the Schr\"odinger-von Neumann equation,
\begin{equation}
\rho_{t+\Delta t} = e^{\frac{-i\tilde{H}t}{\hbar}}\rho_t e^{\frac{i\tilde{H}t}{\hbar}},
\end{equation}
which was solved numerically using the Armadillo linear algebra library in C++ where the matrix exponentials were solved using scaling methods ($e^A = \prod\limits^s e^{\frac{A}{2^s}}$) and a Taylor expansion. In the experiments, we apply microwave pulses with square envelopes that have a finite rise time due to the limited bandwidth of the I/Q channels of the MW vector source.  For simplicity, we approximate these MW pulses with a perfect square envelope. On the other hand, the detuning pulses were modelled with a finite rise/fall time using a Fermi-Dirac function in order to take (a)diabatic effects into account. The finite rise time was set to 2~ns based on the cut-off frequency of low-pass filter attached to the lines used to pulse  the detuning pulses.

\noindent {\bf Modelling noise in the simulation.} In the model we include three different noise sources. The first two noise sources are from fluctuating nuclear spins in the natural silicon quantum well which generate quasi-static magnetic noise which couples to the qubits via the $Z\otimes I$ and $I \otimes Z$ terms in the Hamiltonian. These fluctuations are treated as two independent noise sources  as D1 and D2 are in different locations in the quantum well and will sample the field from different nuclear spins.  The third noise source is charge noise which can couple to the qubits via the magnetic field gradient from the micromagnets which we model as magnetic noise on the  $Z\otimes I$ and $I \otimes Z$ terms in the Hamiltonian. In addition, charge noise also couples to the spins via the exchange coupling which leads to noise on the $Z \otimes Z$ term in the Hamiltonian. 

In our simulations, we treat these noise sources as quasistatic where the noise is static within each cycle and only changes between measurement cycles. This approximation is reasonable because the noise in the system is pink, with low frequencies in the power spectrum more pronounced \cite{Kawakami2016}. The static noise due to each noise source was modelled by sampling a random value from a Gaussian distribution with a standard deviation, $\sigma$, corresponding to the contribution to dephasing of that noise process.  After sampling the static noise, the time evolution of the qubits during a gate sequence was calculated. This time evolution was averaged over many repetitions to give the final result where for each repetition new values for the static noise were sampled. In total, for each simulation we performed 5000 repetitions to ensure convergence. 

In the experiment, single-qubit gates are performed at higher detuning near the center of the (1,1) $\epsilon = -3~$meV where the exchange is low, $J_{\mathit{off}} = 270$~kHz, and a two qubit CZ gate is performed by pulsing to low detuning $\epsilon = -0.7$~meV where the exchange is high, $J_{\mathit{on}} = 6$~MHz. To estimate the relative effect of charge noise on the $Z\otimes I$, $I \otimes Z$, and $Z \otimes Z$ terms at these two detuning points, we use the spectroscopy data of the qubits as a function of detuning energy shown in Extended Data Fig.~8. The four observed resonances correspond to the four transitions shown in Extended Data Fig.~8(c) between the $\ket{00}$,$\ket{01}$,$\ket{10}$,$\ket{11}$ eigenstates. From the fits of this data we can estimate the derivative of the transition energy from state $\ket{i}$ to $\ket{j}$  at a particular detuning, $\frac{dE_{\ket{i} \rightarrow \ket{j}}}{d \epsilon}|_\epsilon$, which is directly proportional to the magnitude of fluctuations in the transition energy under the influence of charge noise. Fixing the energy of the $\ket{00}$ state, from these derivatives we can calculate the relative noise levels on the other energy eigenstates,
\begin{equation}
	B(\epsilon) = \begin{pmatrix}
		0 \\
		\frac{\partial E_{\ket{00} \leftrightarrow \ket{01}}}{\partial \epsilon}|_\epsilon \\
		\frac{\partial E_{\ket{00} \leftrightarrow \ket{10}}}{\partial \epsilon}|_\epsilon \\
		\frac{\partial E_{\ket{00} \leftrightarrow \ket{01}}}{\partial \epsilon}|_\epsilon + \frac{\partial E_{\ket{01} \leftrightarrow \ket{11}}}{\partial \epsilon}|_\epsilon
	\end{pmatrix}
\label{noiselevels}	
\end{equation}
In the regime where $J \ll \Delta v$, the Hamiltonian of the system can be approximated as $H = -B_1 (Z \otimes I)  -B_2 (I \otimes Z) + J(Z \otimes Z) - J/4 (I \otimes I)$. The relative noise on $B_1$, $B_2$, and $J$ can be be found by decomposing the four noise levels in Eq.~\ref{noiselevels} in terms of the basis ($-Z\otimes I, -I\otimes Z, Z\otimes Z, -I\otimes I/4$) by calculating $A^{-1}*B(\epsilon)$ where,
\begin{equation}
	A = \begin{pmatrix}
		-1/2 & -1/2 & 1/4 & -1/4 \\
		-1/2 & 1/2 & -1/4 & -1/4 \\
		1/2 & -1/2 & -1/4 & -1/4 \\
		1/2 & 1/2 & 1/4 & -1/4 \\
	\end{pmatrix}
\end{equation}
We estimate the relative composition of the noise for ($B_1$, $B_2$, $J$) at $\epsilon = -3$~meV to be  (0.12, 0.24, 0) and at $\epsilon = -0.7$~meV ($J = 6$~MHz) to be (0.61, 0.23, 0.26). 
Note that this is a crude approximation since we only take into account voltage noise along the detuning axis,whereas in reality charge noise acts also along other axes. Not included in the simulation are calibration errors. Based on the the AllXY and Ramsey calibration experiments (see Supplementary S2), few $\%$ miscalibrations are possible.

\noindent {\bf Estimating charge noise from the decay of the decoupled CZ oscillations.} Dephasing due to charge noise coupling into the double dot system via the exchange energy is measured by varying the duration of the decoupled CZ gate between two $\pi/2$ pulses on Q1 as shown in Extended Data Fig.~9 for $J = 6~MHz$. The decoupled CZ gate removes the effect of quasi-static noise on the $Z\otimes I$ and $I\otimes Z$ terms in the Hamiltonian  and the decay of the oscillations $T_2 = 1640$~ns is assumed to be due to noise on the $Z\otimes Z$ term. The data is fitted using either a Gaussian (black line) or exponential decay (red line). The exponential decay seems to fit best to the data which suggests that either higher frequency noise plays a role \cite{Dial2013} or the origin of the noise is from a few two-level fluctuators \cite{Ithier2005}. Since the decoupling CZ decay is slower than the not-decoupled CZ decay, there is also a significant quasi-static noise contribution. For simplicity, we only include the quasi-static contribution in our noise model. For Gaussian quasi-static noise with a standard deviation $\sigma_\epsilon$, the decay time is,
\begin{eqnarray}
1/T_2  = \frac{1}{2} \frac{\partial J}{\partial \epsilon}|_\epsilon \frac{\sigma_\epsilon}{\sqrt{2}\hbar}
\end{eqnarray}
The factor of $\frac{1}{2}$ is needed as it is the noise on $J/2$ which contributes to the decay. This is because the target qubit precesses with frequency of $J/2$ (ignoring the $I\otimes Z$ and $Z\otimes I$ terms) when the control qubit is in an eigenstate. From the dephasing time and $\frac{\partial J}{\partial \epsilon}|_\epsilon = 1.0\times10^{-4}$ extracted from Extended Data Fig.~8(a-b) we can estimate the charge noise on detuning to be $11~\mu$eV.  The data in Extended Data Fig.~9 used to extract this value of charge noise was taken over $\sim 40$~minutes with no active calibration on the detuning pulse. The time needed for each single-shot measurement was $\sim 10$~ms. 

\noindent{\bf Simulations of the two qubit algorithms.} To describe the double dot system used in the experiment, we used the following parameters in the Hamiltonian. The qubit frequencies were chosen to be $B_1 = 18.4$~GHz,  $B_2 = 19.7$~GHz, and the on-site charging energies to be $U_1 = U_2 = 3.5$~meV, comparable to the experimental values. The tunnel coupling was chosen to be $t = 210$~MHz so that the residual exchange energy $J_{\mathit{off}}$ was equal to $300$~kHz, giving a similar $J_{\mathit{off}}$  as measured in the experiment. The two-qubit gates are implemented by choosing a value of $\epsilon$ where $J=6$~MHz, when diagonalizing the Hamiltonian $\hat{H}$. 

The results of the simulations for the Deutsch-Josza algorithm and the Grover algorithm using both the CZ gate and the decoupled CZ gate are shown in Fig.~3 and Extended Data Fig.~10. The amplitudes for the three noise sources used in the simulations were identical for all 16 panels. The value of charge noise used was 11~$\mu$eV (see above) while the nuclear spin noise for Q1 and Q2 was chosen to give the single qubit decoherence times $T_2^* = 1000$~ns and  $T_2^* = 600$~ns  measured in the Ramsey experiment in the Extended Data Fig.~3. This gave a dephasing time of Q1 (Q2) due to nuclear spin of $T_{2nuc}^*= 1200$~ns (800~ns). The simulations reproduce many of the features found in the experimental data for the algorithms. 

By simulating the algorithms, we learn that the residual exchange coupling $J_{\mathit{off}}$ during single-qubit gates has little effect ($<2\%$) on the result of the algorithms. Furthermore, we find that without noise on the single-qubit terms, it is difficult to get a consistent agreement with the data. Additional noise on the coupling strength improves the agreement.  Different from the cases of the Deutsch-Jozsa algorithm and the conventional Grover algorithm, the simulation for the decoupled version of Grover’s algorithm predicts a better outcome than the experiment. This case uses the longest sequence of operations, leaving most room for discrepancies between model and experiment to build up. Those could have a number of origins: (i) the implementation of the static noise model is not accurate enough, (ii) non-static noise plays a role, (iii) the calibration errors in the gates that were left out of the simulation, and (iv) variations in the qubit parameters and noise levels between experiments. Finally, we note that initialisation and readout errors are not taken into account in the simulations. Since initialisation errors are negligible and the data shown was renormalised to remove the effect of readout errors, the simulated and experimental results can be compared directly.

\noindent{\bf Data availability.} The raw data and analysis that support the findings of this study are available in the Zenodo repository (https://doi.org/10.5281/zenodo.1135014).

\begin{figure*}[ht]
\includegraphics[scale=0.7]{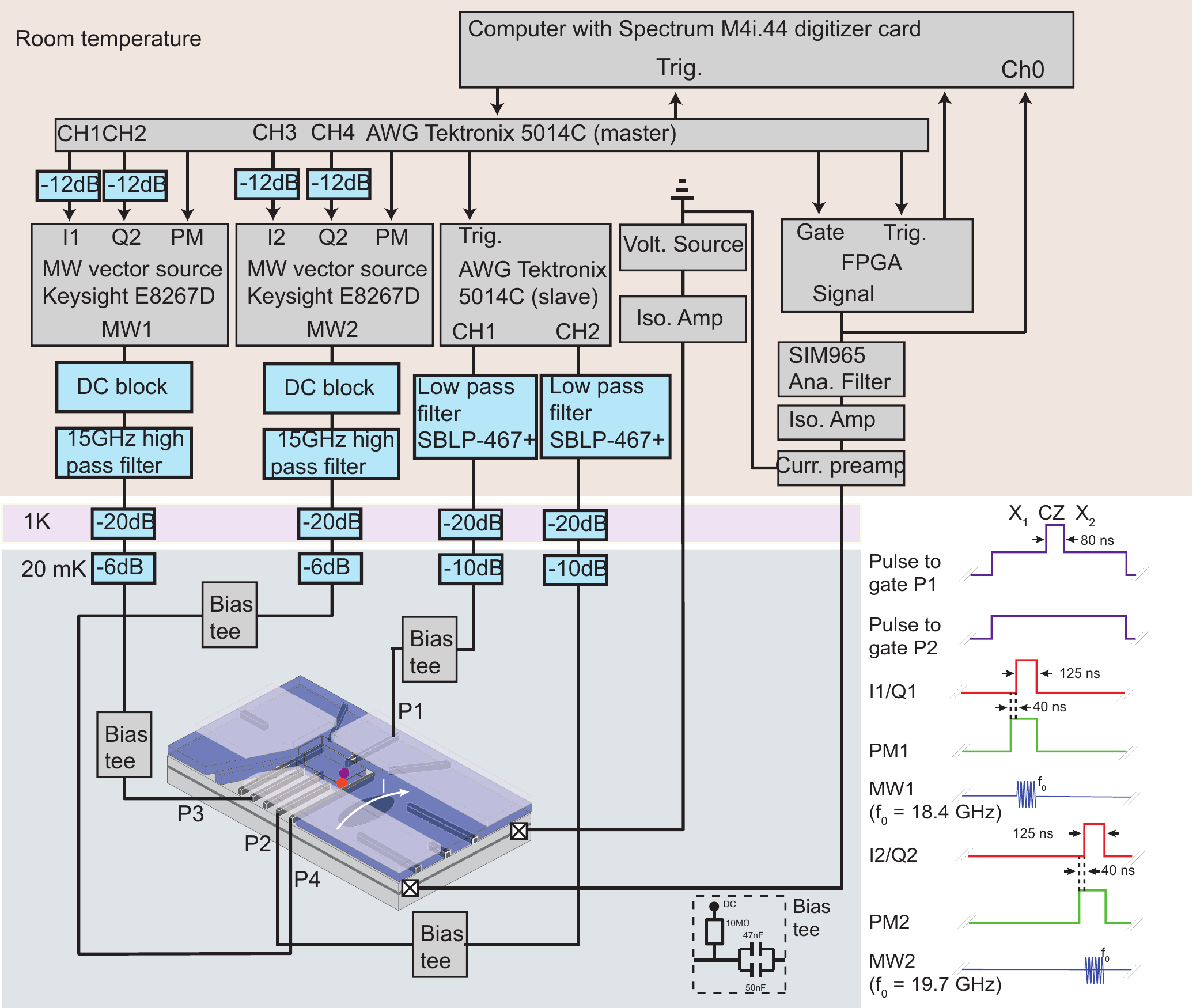}
\caption{{\bf Schematic of the measurement setup.} The sample was bonded to a printed circuit board (PCB)  mounted onto the mixing chamber of a dilution refrigerator. All measurements were performed at the base temperature of the fridge, $T_{base} \sim 20$~mK. DC voltages are applied to all the gate electrodes using room temperature (RT) DACs via filtered lines (not shown). Voltage pulses are applied to plunger gates P1 and P2 using a Tektronix 5014C arbitrary waveform generator (AWG) with 1~GHz clock rate. The signals from the AWG's pass through a RT low-pass filter  and attenuators at different stages of the fridge and are added to the DC signals via bias tees mounted on the PCB. Two Keysight E8267D vector microwave sources, MW1 and MW2, are used to apply microwaves ($18 - 20$~GHz) to perform EDSR on Q1 and Q2, respectively. The signals pass through RT DC blocks, homemade 15~GHz high-pass filters, and attenuators at different stages of the fridge and are added to the DC signals via bias tees mounted on the PCB. The output of the MW source (phase, frequency, amplitude, duration) is controlled with  I/Q vector modulation. The I/Q signals are generated with another Tektronix 5041C which is the master device for the entire setup and provides trigger signals for the other devices. In addition to the vector modulation we employ pulse modulation to give an on/off microwave power output ratio of ~120~dB. While I/Q modulation can be used to output multiple frequencies, the bandwidth of the AWG was not enough to control both qubits with one microwave source due to their large separation in frequency (1.3~GHz). The sensor current, $I$, is converted to a voltage signal with a homebuilt preamplifier and an isolation amplifier is used to separate the signal ground with the measurement equipment ground to reduce interference. Following this, a 20~kHz Bessel low-pass filter is applied to the signal using a SIM965 analog filter. An FPGA analyses the voltage signal during the readout and assigns the trace to be spin-up if the voltage falls below a certain threshold. The  voltage signal can also be measured with a digitizer card in the computer. The shape of the pulses generated by the AWGs and MW sources during qubit manipulation with the typical timescales is shown in the lower left. Square pulses were used to perform the CZ gate and as the input for the I/Q modulation to generate MW pulses. The pulse modulation was turned on 40~ns before turning on the I/Q signal due to the time needed for the modulation to switch on.}
\label{Extendedfigure1}
\end{figure*}

\begin{figure*}[ht]
\includegraphics[scale=0.80]{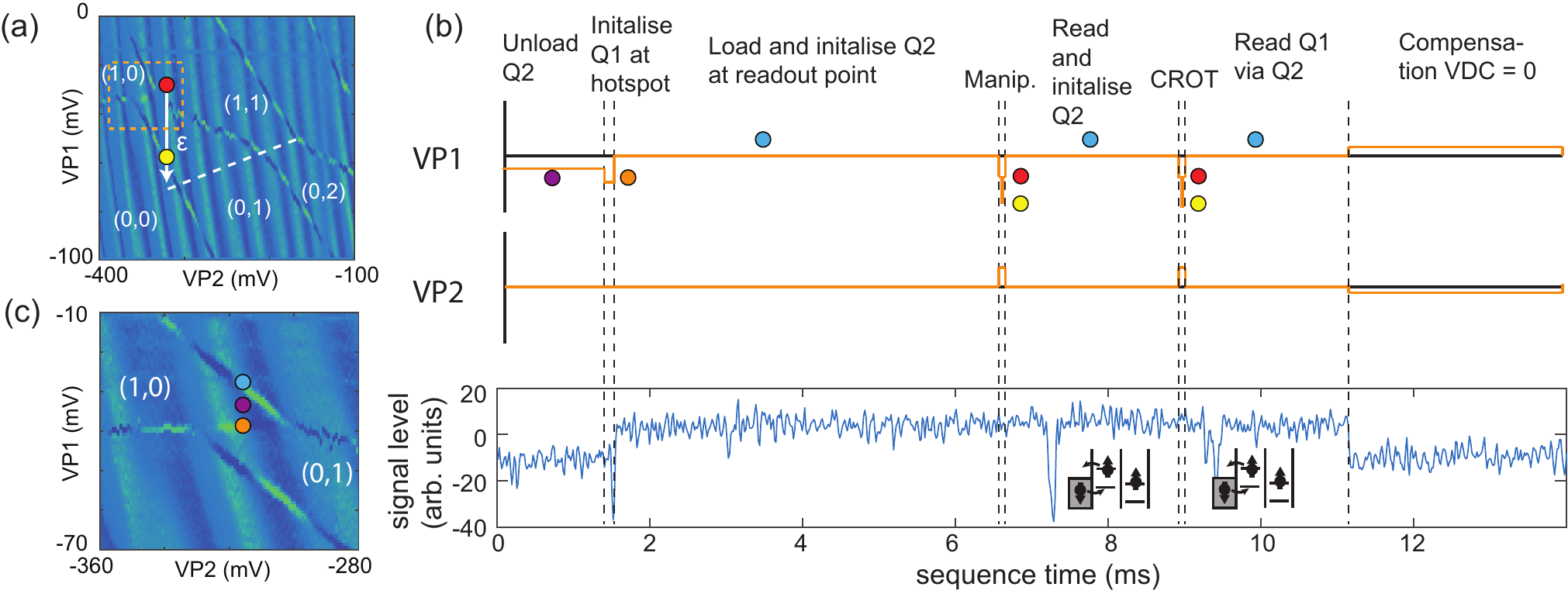}
\caption{{\bf Measurement protocol for two electron spins.} (a) Stability diagram of the double quantum dot showing the positions in gate space used to perform single qubit gates (red circle) and the two-qubit gates (yellow circle). The white dashed line is the (1,1)-(0,2) inter-dot transition line.  The white arrow indicates the detuning axis, $\epsilon$, used in the experiments. Although the detuning pulse for the two-qubit gate crosses the charge addition lines of D1 and D2, the quantum dots remain in the (1,1) charge state as the pulse time is much shorter than the electron tunnel times to the reservoirs.  (b) Plot of the voltage pulses applied to plunger gates P1 and P2 and the response of the quantum dot charge sensor over one measurement cycle. Firstly, D2 is unloaded by pulsing into the (1,0) charge region for 1.5~ms  (purple circle). The electron on D1 is initialised to spin-down by pulsing to a spin relaxation hotspot at the (1,0) and (0,1) charge degeneracy (orange circle) for $50~\mu$s (see Extended Data Fig.~5). D2 is loaded with a spin-down electron by pulsing to the readout position for 4~ms  (blue circle). During manipulation, the voltages on the plunger gates are pulsed to the red circle for single-qubit gates and to the yellow circle for two qubit gates where the exchange is $\sim 6$~MHz. After manipulation, the spin of the electron on D2 is measured by pulsing to the readout position (blue circle) for $0.7$~ms where the Fermi level of the reservoir is between the spin-up and spin-down electrochemical potentials of D2. If the electron is spin-up it can tunnel out followed by a spin-down electron tunnelling back in. These two tunnel events are detected by the QD sensor as a single blip in the current signal. An additional $1.3$~ms  is spent at the readout position so that D2 is initialised to spin-down with high fidelity. Following this, Q1 is measured by first performing a CROT at the yellow circle so that $\alpha \ket{00} + \beta \ket{10} \xrightarrow{CROT12} \alpha \ket{00} + \beta \ket{11}$. A projective measurement of Q1 is then performed by measuring Q2 at the readout position for $0.7$~ms (blue circle). Finally, we add a compensation pulse to VP1 and VP2 so that over the measurement cycle $V_{DC} = 0$ to mitigate charging effects in the bias tees. (b) Close-up of the stability diagram in (a) showing the positions in gate-space used for initialisation and readout.}
\label{Extendedfigure2}
\end{figure*}

\begin{figure*}[ht]
\includegraphics[scale=0.80]{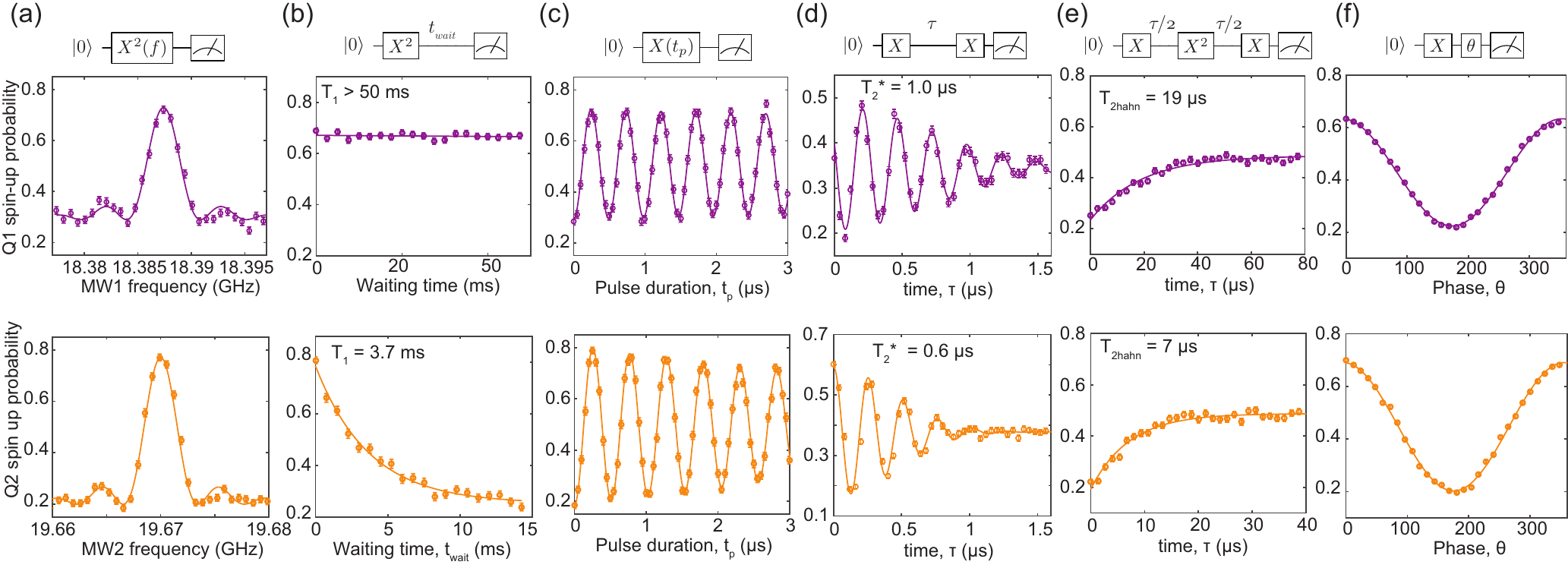}
\caption{{\bf Single qubit properties and two-axis control.} The purple (top) and orange (bottom) data correspond to measurements performed on Q1 and Q2, respectively, in the (1,1) regime (red circle in Extended Data Figure~2). (a) Spin-up fraction as a function of the MW frequency of an applied $\pi$ pulse showing a resonant frequency of 18.424~GHz (19.717~GHz) for Q1 (Q2). (b) The spin relaxation time is measured by preparing the qubit to spin-up and varying the wait time before readout. From the exponential decay in the spin-up probability we measure $T_1>50$~ms ($T_1=3.7 \pm 0.5$~ms) for Q1 (Q2). (c) Spin-up probability as a function of MW duration showing Rabi oscillations of 2.5~MHz for Q1 and Q2. (d) The dephasing time is measured by applying a Ramsey pulse sequence and varying the free evolution time, $\tau$. Oscillations were added artificially to help fit of the decay by making the phase of the last microwave pulse dependent on the free evolution time, $\phi  = \sin(\omega \tau)$ where $\omega = 4$~MHz. By fitting the data with a Gaussian decay, , $P_{\ket{1}} \propto \exp{[-(\tau/T_2^*)^2]} \sin(\omega \tau)$, we extract $T_2^* = 1.0\pm 0.1~\mu$s ($T_2^* = 0.6 \pm 0.1~\mu$s) for Q1 (Q2). In the measurement for Q1 the first $\pi/2$ MW pulse is a Y gate. The Ramsey measurement was performed over $\sim$20~mins with the frequency calibrated every $\sim$1~min. (e) The coherence time of Q1 (Q2) can be extended  to $T_{2Hahn} = 19\pm 3~\mu$s ($7\pm1$~$\mu$s) by a Hahn echo sequence. The coherence time is extracted from an exponential fit to the spin-up probability as a function of the free evolution time in the Hahn echo sequence. (f) Full two axis control is demonstrated by applying two $\pi/2$ pulses and varying the phase of the last $\pi/2$ pulse.}
\label{Extendedfigure3}
\end{figure*}

\begin{figure}[ht]
\includegraphics[scale=1.0]{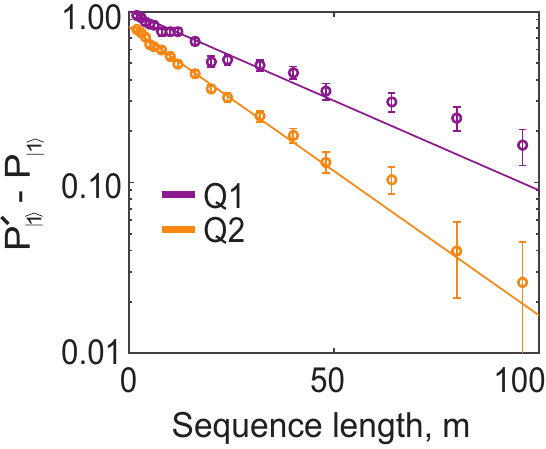}
\caption{{\bf  Randomised benchmarking of single-qubit gates.} Randomised benchmarking of the single qubit gates for each qubit is performed by applying a randomised sequence of a varying number of  Clifford gates, $m$, to either the $\ket{1}$ or $\ket{0}$ state and measuring the final spin-up probability $P_{\ket{1}}'$ or $P_{\ket{1}}$, respectively. All gates in the Clifford group are decomposed into gates from the set $\{I, \pm X, \pm X^2, \pm Y, \pm Y^2\}$. The purple (orange) data points show the difference in the spin-up probabilities $P_{\ket{1}}' - P_{\ket{1}}$ for Q1 (Q2) as a function of sequence length. For each sequence length, $m$, we average over 32 different randomised sequences. From an exponential fit (solid lines) of the data, $P_{\ket{1}}' - P_{\ket{1}}'= ap^m$, we estimate an average Clifford gate fidelity $F_C = 1 - (1-p)/2$ of 98.8$\%$ and 98.0$\%$ for Q1 and Q2, respectively. The last three data points from both data sets were omitted from the fits as they begin to deviate from a single exponential\cite{Kawakami2016}. All errors are $1\sigma$ from the mean.}
\label{Extendedfigure4}
\end{figure}

\begin{figure}[ht]
\includegraphics[scale=0.90]{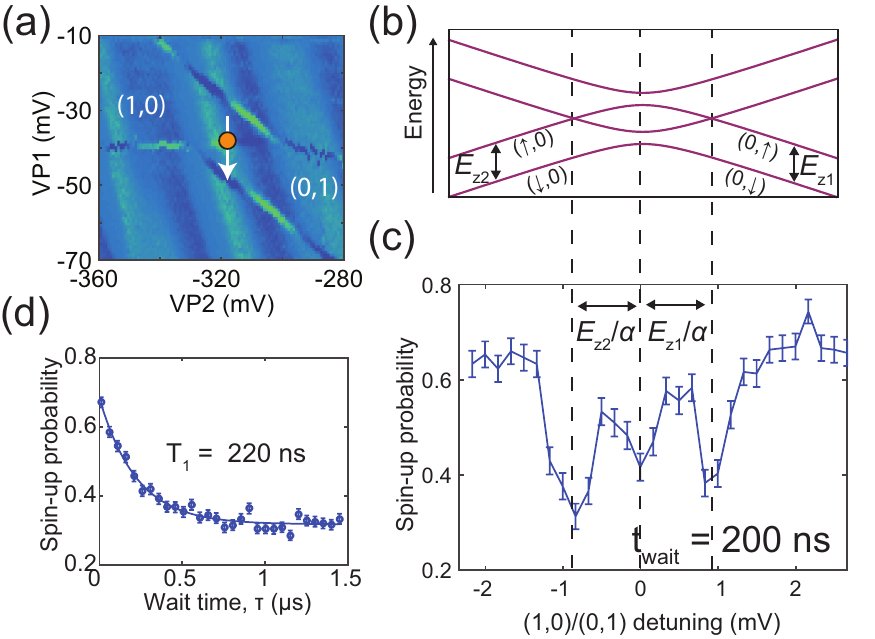}
\caption{{\bf Spin relaxation hotspots used for high fidelity initialisation.} (a) Close-up stability diagram of the (1,0) to (0,1) charge transition. The white arrow defines the detuning axis between D1 and D2 controlled with P1. (b) Schematic of the energy level diagram as a function of detuning for one electron spin in a double quantum dot. (c) Spin relaxation hotspots are measured by first preparing the electron on D1 to spin-up using EDSR, applying a voltage pulse along the detuning axis (white arrow in (a)) for a wait time of $200$~ns, and performing readout of the electron spin. We observe three dips in the spin-up probability corresponding to spin relaxation hot spots. The first and third hotspot are due to anticrossings between the $(0, \downarrow)$ and $(\uparrow, 0)$ states and the $(\downarrow, 0)$ and $(0, \uparrow)$ states \cite{Srinivasa2013}. The second hotspot occurs at zero detuning.  The voltage separation between the first and third hot spot corresponds to the sum of the Zeeman energy of D1 and D2 divided by the gate lever arm $\alpha$ along the detuning axis. Knowing precisely the Zeeman energies from EDSR spectroscopy we can accurately extract the gate lever arm  to be $\alpha = 0.09$e. (d) The spin relaxation time at zero detuning (orange circle in (a)) is found to be $T_1 = 220$~ns by measuring the exponential decay of the spin-up probability as a function of wait time, $\tau$, at zero detuning.}
\label{Extendedfigure5}
\end{figure}

\begin{figure}[ht]
\includegraphics[scale=0.90]{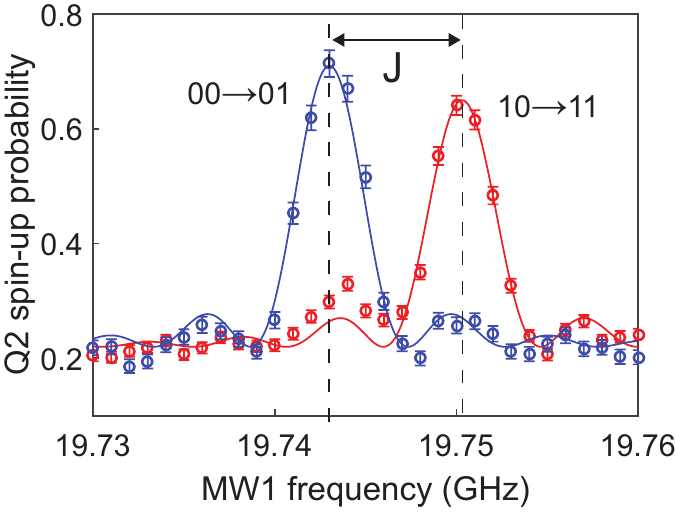}
\caption{{\bf Two-qubit controlled rotation (CROT) gate.} (a) Microwave spectroscopy of Q2 close to zero detuning between the (1,1) and (0,2) state (yellow dot in Extended Data Fig.~2(a)) where the exchange coupling is on. The blue and red curve show the resonance of Q2 after preparing Q1 into spin-down  or up, respectively. The resonance frequency of Q2 shifts by the exchange coupling and by applying a $\pi$ pulse at one of these frequencies we can perform a CROT, which is equivalent to a CNOT up to a $\hat{z}$ rotation. As discussed in the main text, this CROT gate is used to perform the projective measurement of Q1.}
\label{Extendedfigure6}
\end{figure}

\begin{figure}[ht]
\includegraphics[scale=0.90]{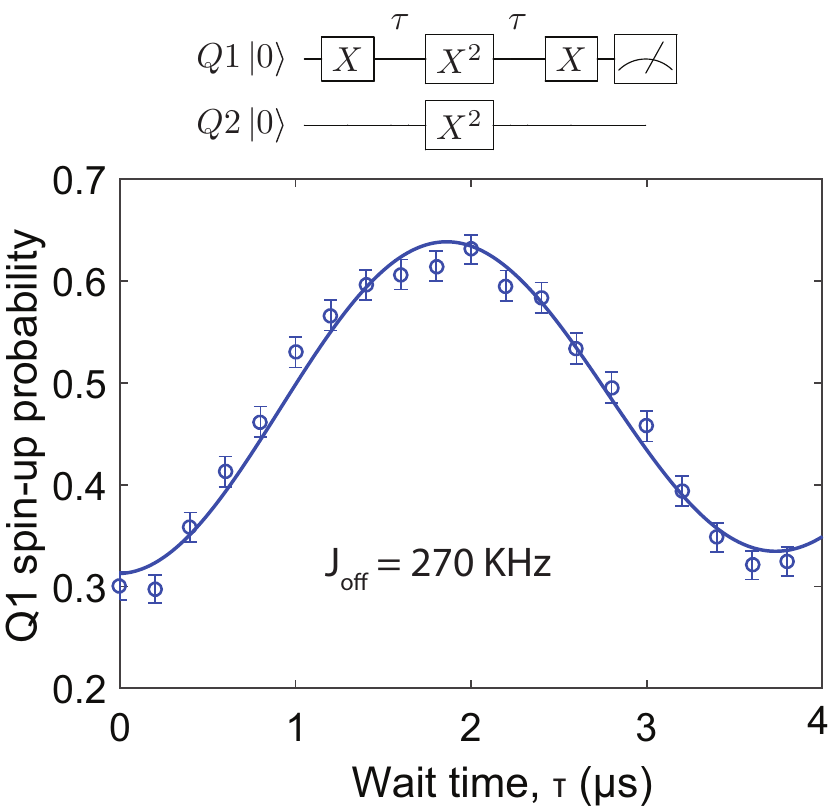}
\caption{{\bf Measurement of $J_{\mathit{off}}$ using a decoupling sequence.} The exchange coupling $J_{\mathit{off}}$ during single-qubit gates is measured using a two-qubit Hahn echo sequence which cancels out any unconditional $\hat{z}$ rotations during the free evolution time $\tau$. Fitting the spin-up probability as a function of free evolution time $\tau$ using the functional form $\mathrm{sin}(2\pi J_{\mathit{off}}\tau)$, we extract $J_{\mathit{off}} = 270$~kHz.}
\label{Extendedfigure7}
\end{figure}

\begin{figure*}[ht]
	\includegraphics[width = \linewidth]{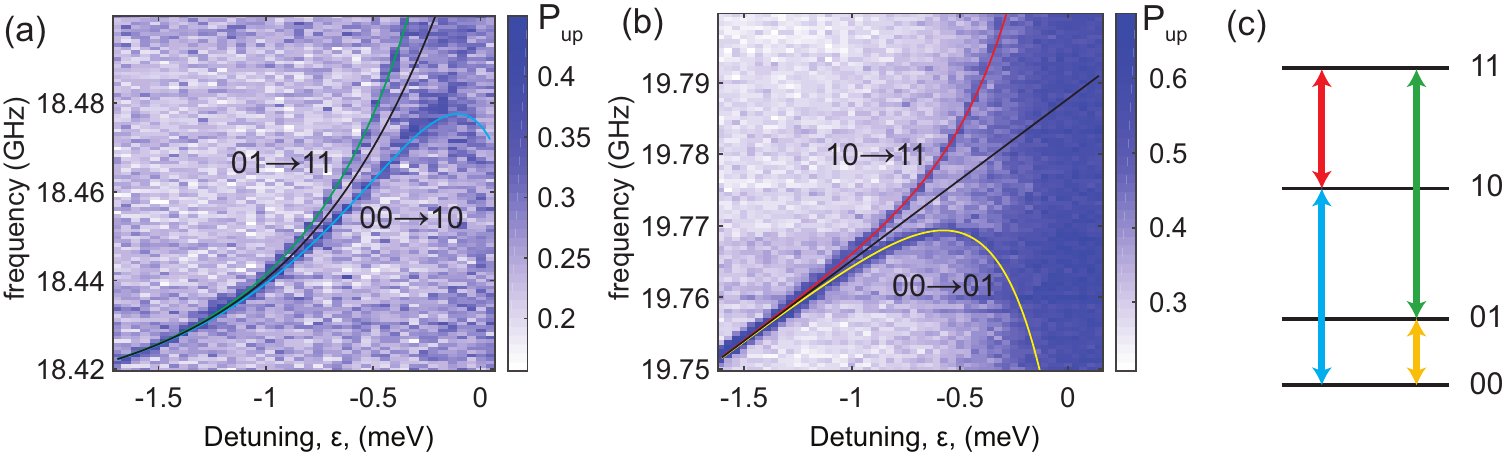}
	\caption{{\bf Microwave spectroscopy of Q1 and Q2.} (a,b) Spectroscopy of (a) Q1 and (b) Q2 versus detuning energy, $\epsilon$, after initialising the other qubit to $(\ket{0} + \ket{1})/\sqrt{2}$. Towards $\epsilon =0$  there are two resonances for Q1 (Q2) which are separated by the exchange energy, $J(\epsilon)/h$. As discussed in the manuscript, the Zeeman energy $E_{Z}(\epsilon)$ of Q1 and Q2 also depends on detuning as changes to the applied voltages will shift the position of the electron in the magnetic field gradient.	The four resonance frequencies are fitted (green, blue, red and yellow lines) with $f_{jk} = E_{Zj}(\epsilon) + (-1)^{k+1} J(\epsilon)$ where $j$ denotes the qubit and $k$ denotes the state of the other qubit. The data is fit well using $J(\epsilon) \propto e^{c_1\epsilon}$, $E_{Z1}(\epsilon) \propto e^{c_2\epsilon}$, and   $E_{Z2}(\epsilon) \propto \epsilon$.  The fitted Zeeman energies of Q1 and Q2 are shown by the black lines. We observe that the Zeeman energy of Q1 has an exponential dependence towards the (0,2) charge regime ($\epsilon = 0$) which can be explained by the electron delocalising from D1 towards D2 which has a significantly higher Zeeman energy.  (c) Schematic showing the color coded transitions that correspond to the resonances in (a,b).}
	\label{Extendedfigure8}
\end{figure*}

\begin{figure}[ht]
\includegraphics[scale=0.9]{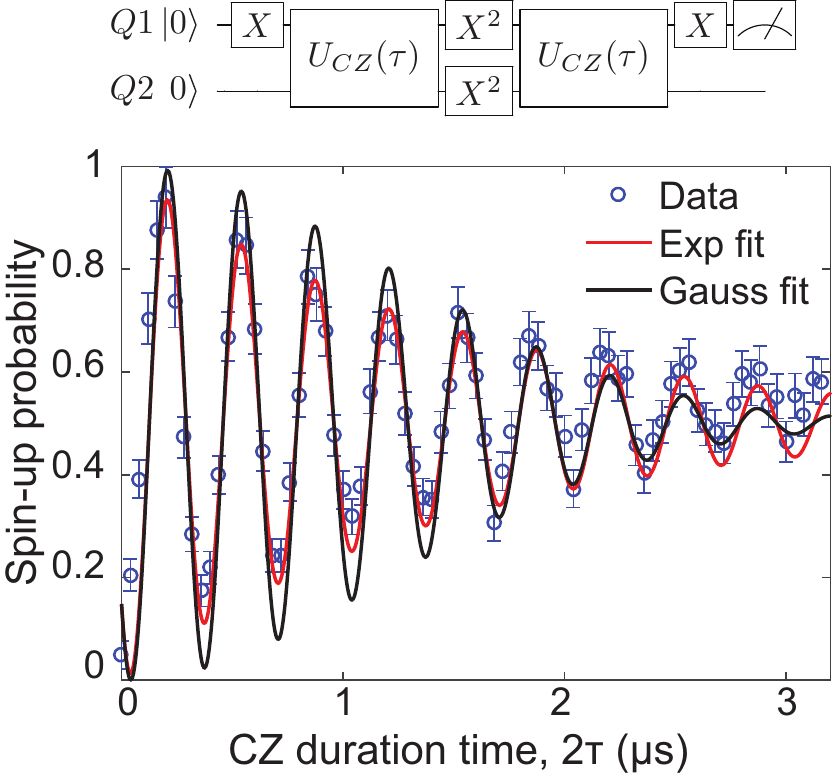}
\caption{{\bf Decay of the decoupled CZ oscillations.} The normalised spin up probability of Q1 as a function of the total duration time, $2\tau$, of the two CZ gates in the decoupled CZ sequence. The data is fitted using a sinusoid, $P_{\ket{1}} = 0.5\sin{2\pi J \tau} +0.5$, with either a Gaussian (black line), $e^{-(2\tau/T_2)^2}$, or exponential (red line), $e^{-2\tau/T_2}$, decay. From these fits we find a decay time of $T_2 = 1.6~\mu$s.}
\label{Extendedfigure9}
\end{figure}

\begin{figure*}[ht]
\includegraphics[scale=0.90]{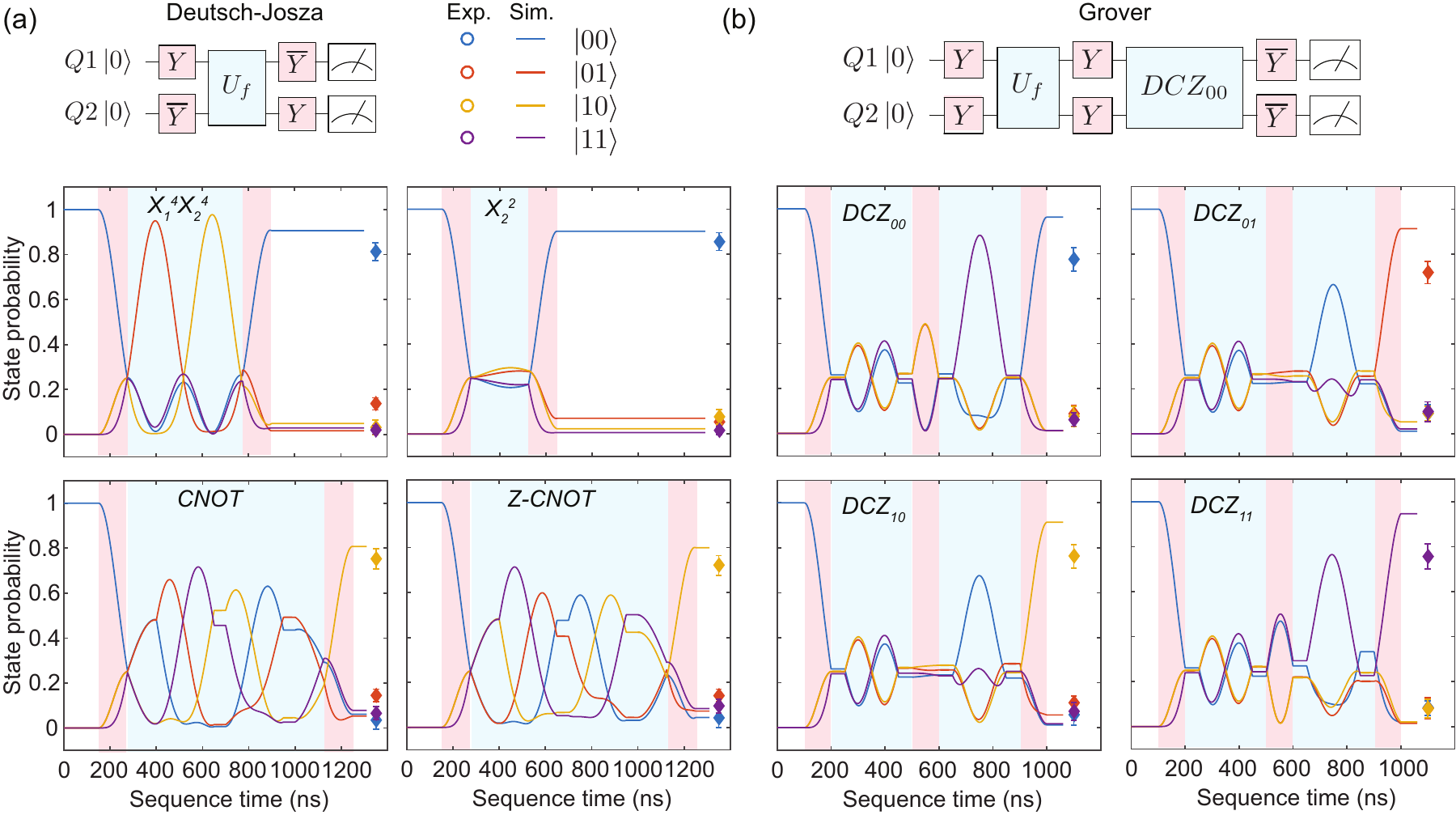}
\caption{{\bf Simulation of the Deutsch-Josza and Grover algorithms using the decoupled CZ gate.} Two-spin probabilities as a function of the sequence time during the (a) Deutsch-Josza algorithm  and the (b) Grover search algorithm for each function using the decoupled version of the two-qubit CZ gate. The solid lines show the outcome of the simulations which include decoherence due to quasi-static charge noise and nuclear spin noise.}
\label{Extendedfigure10}
\end{figure*}

\clearpage
\widetext
\begin{center}
\textbf{\large Supplementary Information: A programmable two-qubit quantum processor in silicon}
\end{center}

\setcounter{equation}{0}
\setcounter{figure}{0}
\setcounter{table}{0}
\setcounter{page}{1}
\makeatletter
\renewcommand{\theequation}{\arabic{equation}}
\renewcommand{\thefigure}{S\arabic{figure}}
\renewcommand{\figurename}{Supplementary Fig.}

\section{S1. Frequency shifts on Q2 due to off-resonant frequency pulses}
\begin{figure}[h]
\includegraphics[scale=0.80]{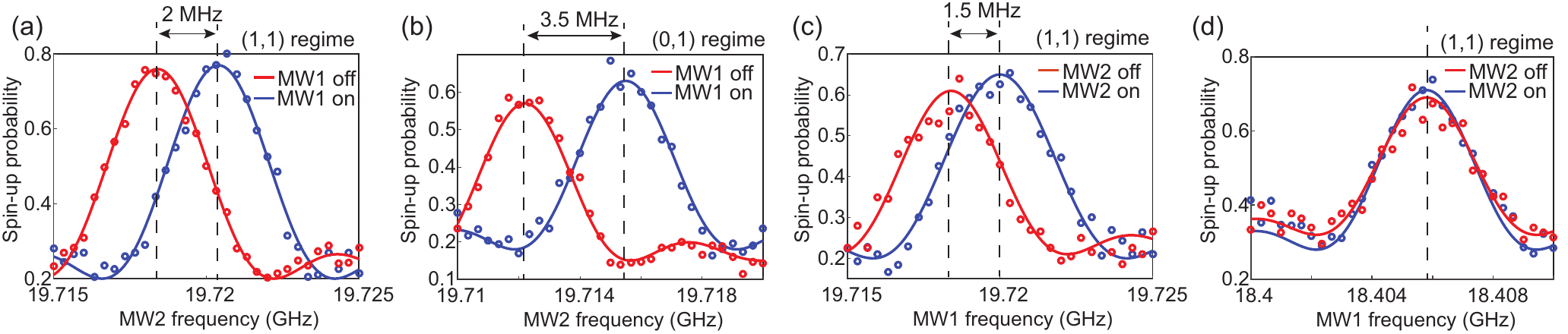}
\caption{{\bf Frequency shift on Q2 due to off-resonant microwaves signals.} (a) Spectroscopy of Q2 with MW2 while MW1 is either off (blue data) or applying off-resonant MWs (red) to plunger gate P3 with a frequency  $f = 18.5$~GHz and power $P = $16~dBm. (b) The same experiment is performed in the (0,1) charge regime where there is only one electron in the double quantum dot. (c) Spectroscopy of Q2 with MW1 while MW2 is off (blue data) or applying off-resonant MWs (red) to plunger gate P4 with a frequency  $f = 18.5$~GHz and power $P = $10~dBm. (d) Spectroscopy of Q1 with MW1 while MW2 is off (blue data) or applying off-resonant MWs (red) to plunger gate P4 with a frequency  $f = 18.5$~GHz and power $P = $22~dBm.  }
\label{Figuresup1}
\end{figure}

\begin{figure}[h]
\includegraphics[scale=0.80]{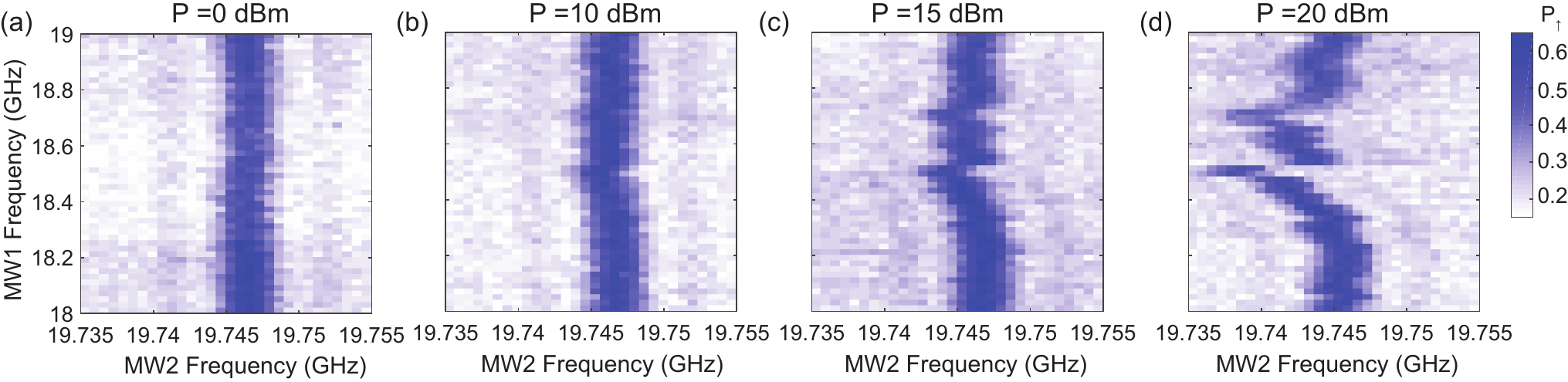}
\caption{{\bf Power dependence of the Q2 frequency shift.} Spectroscopy maps showing the resonance frequency of Q2 measured with MW2 as a function of the off-resonant MW frequency applied using MW1. These maps are measured at MW1 powers (a) 0~dBm, (b) 10~dBm, (c) 15~dBm, and (d) 20~dBm. The larger shifts in the resonance frequency most likely occur at transmission resonances where the power applied on the device is larger. }
\label{Figuresup2}
\end{figure}

As discussed in the main manuscript, we observe a large frequency shift on Q2 while applying off-resonant microwaves (MW). Similar effects have been observed for electron spins bound to single donor atoms \cite{Freer2017} but with a significantly smaller frequency shift and with transient behaviour occurring over $\sim100\mu$s. In our experiment the microwave source MW1 (MW2) applies MWs to gate P3 (P4)  to manipulate Q1 (Q2) as shown in Extended Data Fig.~1. Fig.~\ref{Figuresup1}(a) shows the resonant frequency of Q2 shifting by 2~MHz while off-resonant MWs of 18.5~GHz are applied with MW1 via P3. One possible mechanism to explain this effect is the AC stark shift, where off-resonance MW's will shift the qubit's resonance frequency  ($\omega_L$) by $\sim \omega_R^2/2(\omega_1 - \omega_L)$ away from the drive frequency, $\omega_1$ \cite{Vandersypen2005}. However, this is a negligible effect and the observed frequency shift is towards the off-resonant MW frequency ruling out the AC-stark shift as a possible cause.  We also performed the same experiment in the (0,1) charge regime where we observed similar behaviour (Fig.~\ref{Figuresup1}(b)) eliminating effects due to the coupling between the two electron spins. Fig.~\ref{Figuresup1}(c) shows that the resonant frequency also shifts if instead we apply off-resonant MWs using MW2 via P4 demonstrating that this effect does not depend on the gate electrode/coaxial line used to apply the MW's. Interestingly, we do not see the effect on the other qubit as shown in Fig.~\ref{Figuresup1}(d) where we apply off-resonant microwaves at 18.5~GHz at nearly the maximum output power ($P = 22$~dBm) of the MW2 source suggesting that the effect is due to some property of the quantum dot. The frequency shift is also measured in a Ramsey sequence where during the $\pi/2$ pulse both MW sources are on and during the wait time both MW sources are off. This indicates the frequency shift occurs faster than the Ramsey wait time ($<$100~ns) ruling out local heating effects which would require time to dissipate. Finally, we observe that the frequency shift is strongly dependent on the power of the off-resonance MW's as shown in Fig.~\ref{Figuresup2}.
 
The dependence on the quantum dot properties and power would be compatible with the rectification of the AC signal as an explanation. An asymmetric quantum dot potential will lead to a DC displacement in response to an AC excitation on the gate. We tried to estimate this by measuring the resonance frequency of Q2 as a function of the voltage applied on plunger P3 around the position in gate-space where we perform the single-qubit gates. Over the estimated range of the AC signal, $V_{RMS} \sim 5$~mV for an output power of $P = $16~dBm and measured attenuation of the coaxial line ($\sim 43$~dB at 20~GHz), we observe a change in frequency of $\sim$1~MHz and no measurable non-linearity in the resonance frequency. While this suggests rectification effects are small, it is difficult to get an accurate estimation on the AC signal at the sample and further work is required to rule out this possibility.

\section{S2. Calibration of single-qubit gates.}
To perform accurate single-qubit gates on Q1 and Q2 we need to calibrate the following parameters of our MW pulses, (i) the frequency needed to be on resonance with Q1 and Q2, (ii) the power needed to perform a $\pi/2$ pulse on Q1 and Q2, (iii) the power needed for the 30~MHz off-resonance pulse during Q1 idle times to compensate for the Q2 frequency shift described above.  The resonance frequency of Q1 (Q2) was measured using the Ramsey sequence shown in Supplementary Fig.~\ref{Figuresup_allxy}(a) and corresponds to the MW frequency that gives the maximum spin-up probability for Q1 (Q2).  During the 300~ns wait time we apply the off-resonant MW pulse to Q1 to keep the resonance frequency of Q2 constant throughout the sequence. The power of the applied MW pulses was calibrated using the AllXY calibration sequence. In this sequence, two single-qubit gates, $A$ and $B$ where $A, B \in \{I, X, X^2, Y, Y^2\}$, are applied sequentially to the qubit as shown in Supplementary Fig.~\ref{Figuresup_allxy}(b). All possible combinations of $A$ and $B$ are applied and the final spin-up probabilities are measured. If the gates are ideal then the different combinations  give the expected final probability of either 0, 0.5, or 1. MW power and frequency errors during the single-qubit gates result in characteristic deviations from these probabilities and can be corrected.  In addition, if there is an error in the applied power of the off-resonant MW pulse during the Ramsey calibration so that the resonance frequency of Q2 is not the same during the wait time and the $X$ gates, this will show up as a frequency error in the AllXY sequence and can also be easily corrected. Supplementary Fig.~\ref{Figuresup_allxy}(c) shows an example of the result of the Ramsey and AllXY sequences after all the parameters have been calibrated.

 \begin{figure}[ht]
\includegraphics[scale=0.70]{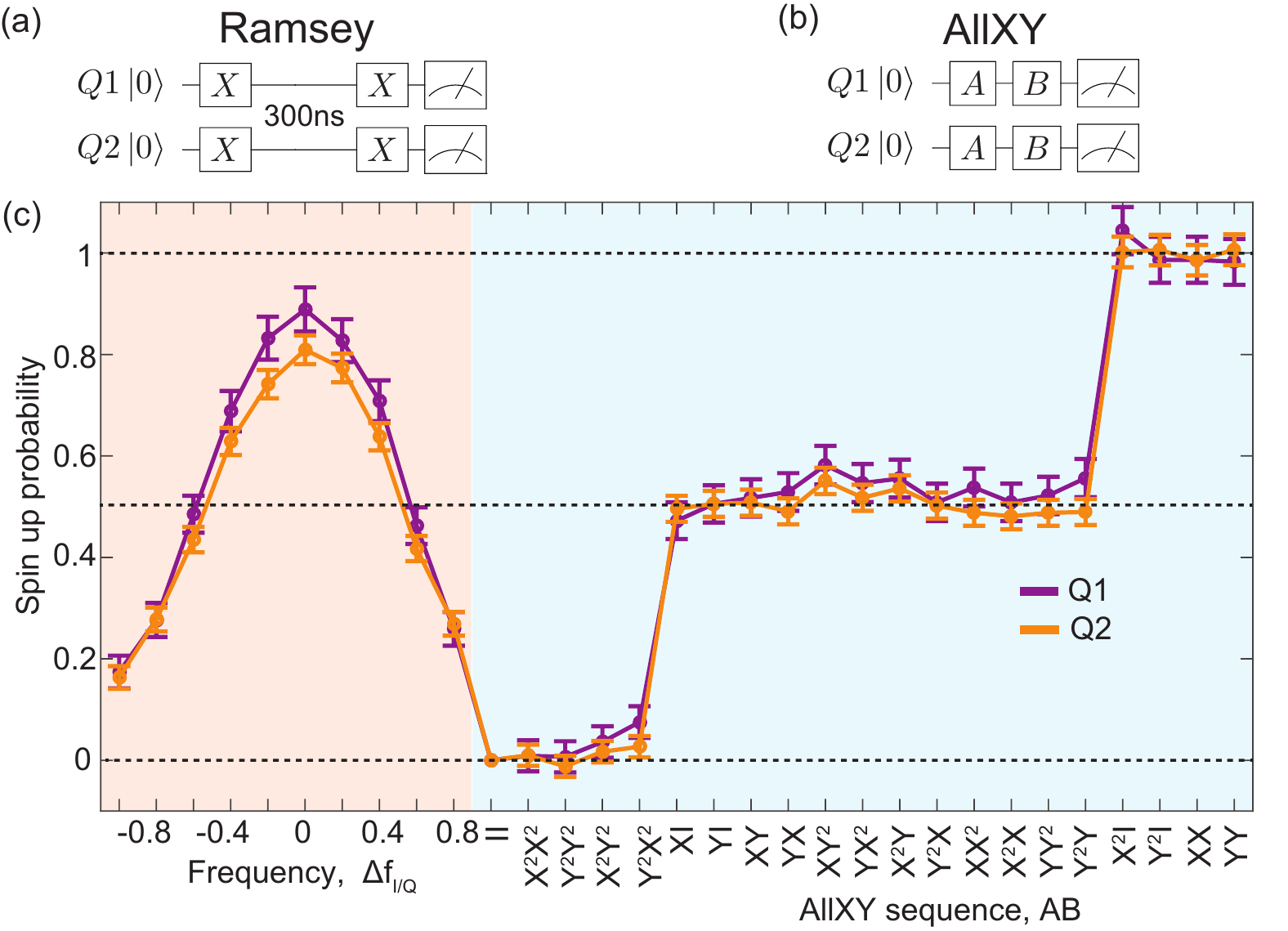}
\caption{{\bf Calibration of the single-qubits gates.} (a) Ramsey and (b) AllXY sequences used to calibrate the single-qubit gates for Q1 and Q2. (c) The measured spin-up probability of Q1 (purple data points) and Q2 (orange data points) during the Ramsey and AllXY experiments after the single-qubit gates have been calibrated. Here, errors due to readout have been removed from the spin-up probabilities. In the Ramsey experiment the MW frequency is swept around the local oscillator frequency of the MW source using I/Q modulation. In the AllXY experiment the x-axis corresponds to the 21 different combinations of the $A$ and $B$ single-qubit gates.}
\label{Figuresup_allxy}
\end{figure}

\section{S3. State tomography of Bell states.}
\begin{figure}[ht]
\includegraphics[scale=0.90]{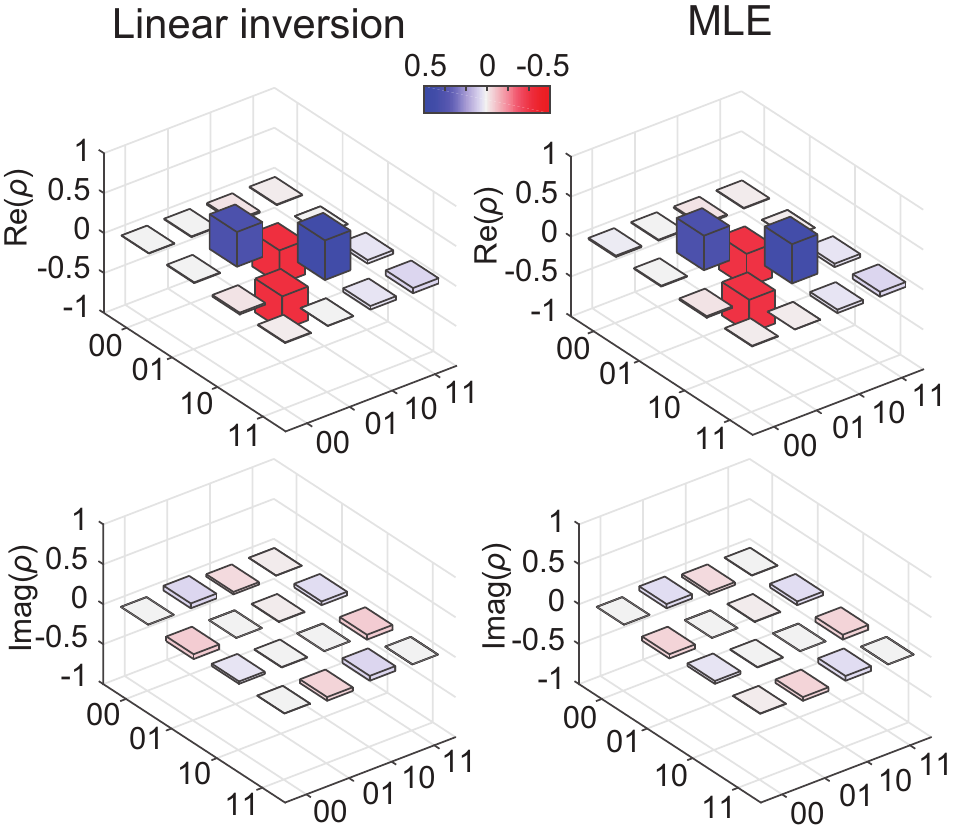}
\caption{{\bf Comparison between the maximum likelihood estimation and linear inversion for the Bell state $\Psi^- = 1/\sqrt{2} (\ket{01} - \ket{10})$.} }
\label{Figuresup3}
\end{figure}

The density matrix of a two-qubit state can be expressed as $\rho = \sum \limits^{16}_{i=1} c_i M_i$ where $M_i$ are 16 linearly independent measurement operators. The coefficients $c_i$ were calculated from the expectation values of the measurement operators either through linear inversion or a maximum likelihood estimation where the later ensures a physical density matrix that is Hermitian and positive semi-definite \cite{James2001}. Fig.~\ref{Figuresup3} shows a comparison of the density matrices for the state $\Psi^- =  (\ket{01} - \ket{10})/\sqrt{2}$ calculated using either linear inversion and MLE. The results are nearly identical indicating the estimated expectation values from the MLE are close to the measured expectation values in the experiment. For all  measured states, the elements of the density matrix calculated with either linear inversion and MLE differ on average by $\sim 0.005$. The calculated density matrices for the four Bell states using MLE are:
\begin{eqnarray}
\Psi^+ &=& \begin{pmatrix}
0.019 + 0i & 0.010 - 0.018i & -0.006 - 0.030i & -0.009 +0.016i  \\
0.010 + 0.018i & 0.425 + 0i& 0.422 - 0.026i & -0.014 + 0.078i\\
-0.006 + 0.030i & 0.422 + 0.026i & 0.493 + 0i  & -0.050 + 0.058i\\
-0.009 - 0.016i &-0.014 - 0.078i & -0.050- 0.058i & 0.063 + 0i
\end{pmatrix},\\
\Psi^- &=& \begin{pmatrix}
0.016 + 0i & 0.009 + 0.052i & -0.016 - 0.035i & -0.008 +0.005i  \\
0.009 - 0.052i & 0.429 + 0i& -0.420 - 0.007i & 0 + 0.050i\\
-0.016 + 0.035i & -0.420 + 0.007i & 0.495 + 0i  & 0.040 - 0.062i\\
-0.008 - 0.005i & 0 - 0.050i & 0.040 + 0.062i & 0.060 + 0i
\end{pmatrix},\\
\Phi^+ &=& \begin{pmatrix}
0.501 + 0i & -0.024 +0.023i & 0.002 + 0.021i & 0.370 + 0.013i  \\
-0.024 - 0.023i & 0.019 + 0i& 0.003 - 0.003i & -0.03 - 0.028i\\
-0.002 - 0.021i & 0.003 + 0.003i & 0.013 + 0i  & 0.017 + 0.019i\\
0.370 - 0.013i &-0.031 + 0.028i & 0.017 - 0.019i & 0.467 + 0i
\end{pmatrix},\\
\Phi^- &=& \begin{pmatrix}
0.505 + 0i & 0.010 - 0.047i & -0.019 + 0.015i & -0.407 +0.001i  \\
0.010 + 0.047i & 0.019 + 0i& -0.002 + 0.010i & -0.024 - 0.025i\\
-0.019 - 0.015i & -0.002 - 0.010i & 0.024 + 0i  & 0.040 + 0.039i\\
-0.407 - 0.001i &-0.024 + 0.025i & 0.040 - 0.039i & 0.452 + 0i
\end{pmatrix},
\end{eqnarray}

\begin{figure}[ht]
\includegraphics[scale=0.80]{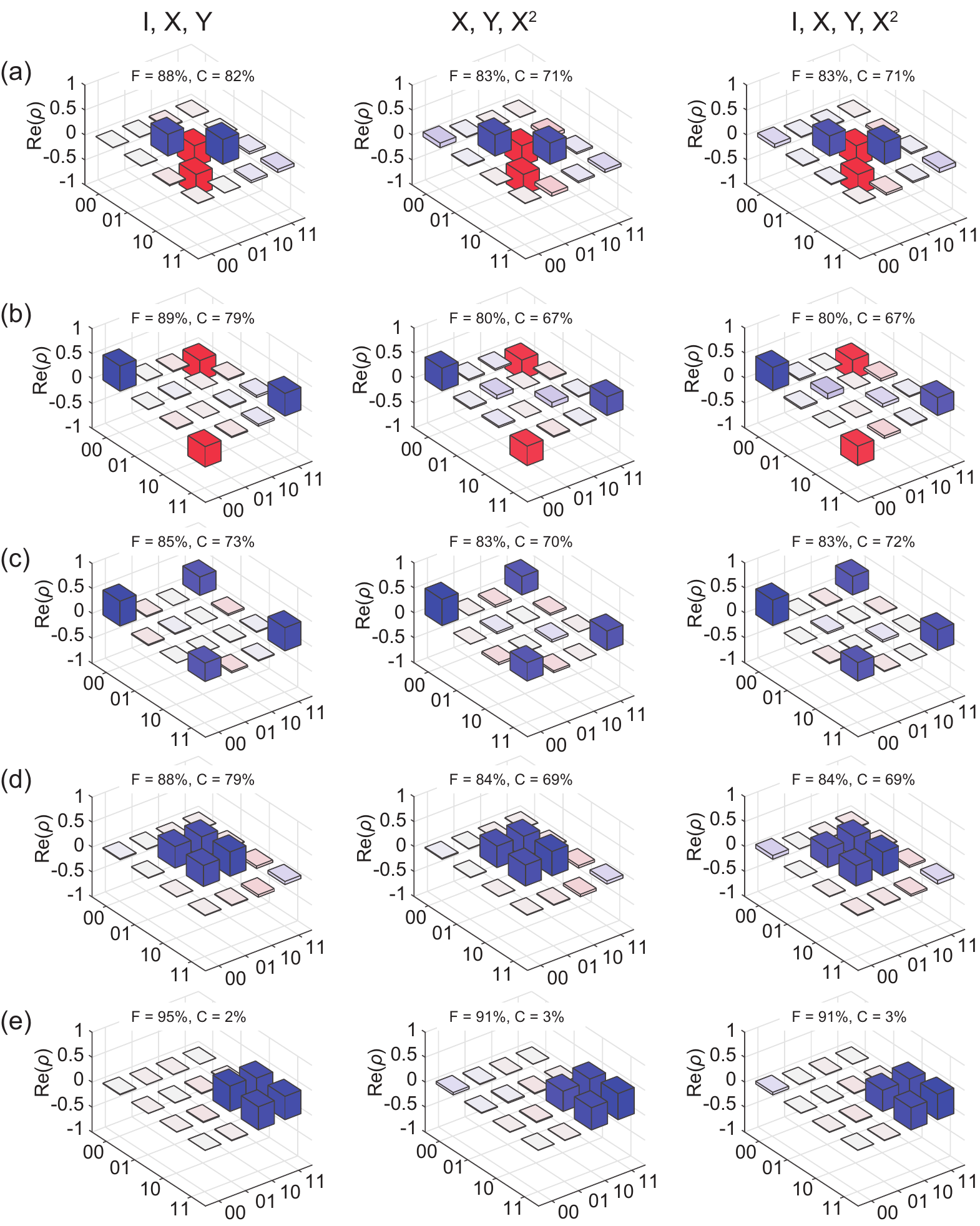}
\caption{{\bf Comparison between density matrices constructed using different prerotations.} The real component of the reconstructed density matrices using a maximum likelihood estimation for the four Bell states (a) $\Psi^+ =  (\ket{01} + \ket{10})/\sqrt{2}$, (b) $\Psi^- =  (\ket{01} - \ket{10})/\sqrt{2}$, (c) $\Phi^+ = (\ket{00} + \ket{11})/\sqrt{2}$, (d) $\Phi^- = (\ket{00} - \ket{11})/\sqrt{2}$, and (e) $\psi  = (\ket{10} + \ket{11})/\sqrt{2}$. Here, the columns label whether the $I,X,Y$ or $X,Y, X^2$ or $I,X,Y,X^2$ prerotations were used to calculate the expectation values in the estimation of the density matrices.}
\label{Figuresup4}
\end{figure}

In quantum state tomography, the density matrix can be reconstructed by measuring the two-spin probabilities after applying 9 combinations of the prerotations $I, X, Y$. In the actual experiment, we also included the prerotation $X^2$ to help detect systematic errors leading to 16 combinations in total. Fig.~\ref{Figuresup4} shows the real component of the estimated density matrices for the four Bell states and $\psi  =  (\ket{10} + \ket{11}/\sqrt{2})$. These were calculated using either all prerotations or a subset of these prerotations, $I,X,Y$ or $X,Y, X^2$. Using either $X,Y, X^2$ or $I,X,Y,X^2$ gives similar results where the state fidelities and concurrences are $2-9\%$ and $4-11\%$ less than those calculated with $I,X,Y$. For the final estimate of the density matrices we use only the prerotations $I,X,Y$ as $I$ should give a better estimate for the expectation values than $X^2$ due to decoherence and small calibration errors in our system.  We did not account for decoherence and other errors in the prerotation pulses, which likely causes us to underestimate the overlaps with the ideal Bell states. Future work will include incorporating the prerotation errors into the state tomography analysis.

\end{document}